\documentclass[preprint]{aastex63}

\usepackage{enumitem,amssymb}
\usepackage{amsmath}
\usepackage{pifont}
\usepackage{multirow}
\received{xxx, 2023}
\revised{yyy, 2024}
\accepted{zzz, 2024}
\submitjournal{PSJ}

\shorttitle{How long-lived grains dominate the shape of the Zodiacal Cloud}
\shortauthors{Pokorn\'{y}, Moorhead, Kuchner, Szalay, Malaspina}
\begin{document}

% \title{Volatile stability and meteoroid bombardment at Ceres/Erosion of volatiles by micrometeroid bombardment on Ceres}
%\title{Collisional Grooming of the Zodiacal Cloud: Jupiter-family comets\\}
\title{How long-lived grains dominate the shape of the Zodiacal Cloud}

\correspondingauthor{Petr Pokorn\'{y}}
\email{petr.pokorny@nasa.gov, pokorny@cua.edu}

\author[0000-0002-5667-9337]{Petr Pokorn\'{y}}
\affiliation{The Catholic University of America, 620 Michigan Ave, NE Washington, DC 20064, USA}
\affiliation{Goddard Space Flight Center, 8800 Greenbelt Rd., Greenbelt, MD, 20771, USA}

\author[0000-0001-5031-6554]{Althea V. Moorhead}
\affiliation{NASA Meteoroid Environment Office, Marshall Space Flight Center EV44, Huntsville, Alabama 35812, USA}

\author[0000-0002-2387-5489]{Marc J. Kuchner}
\affiliation{Goddard Space Flight Center, 8800 Greenbelt Rd., Greenbelt, MD, 20771, USA}

\author[0000-0003-2685-9801]{Jamey R. Szalay}
\affil{Department of Astrophysical Sciences, Princeton University, 4 Ivy Ln., Princeton, NJ 08540, USA}

\author[0000-0003-1191-1558]{David M. Malaspina}
\affiliation{Department of Astrophysical and Planetary Sciences, University of Colorado Boulder, Boulder, CO, USA}
\affiliation{Laboratory for Atmospheric and Space Physics, University of Colorado Boulder, Boulder, CO, USA}


\begin{abstract}
%The Zodiacal cloud, the closest debris cloud to Earth, is known to be continually depleted by particle-particle collisions, particle destruction near the sun, and ejection of particles outside the solar system and at the same time replenished by cometary and asteroid break-ups and activity. Jupiter family-comets are the dominant component of the inner portion of the Zodiacal Cloud in terms of particle mass and cross-section and dominate the mass flux on all terrestrial planets and small bodies in the inner solar system. Here, we tackle the least understood component of Zodiacal Cloud sinks and sources, the particle-particle collisions.

Grain-grain collisions shape the 3-dimensional size and velocity distribution of the inner Zodiacal Cloud and the impact rates of dust on inner planets, yet they remain the least understood sink of zodiacal dust grains. For the first time, we \replaced{use}{combine} the collisional grooming method combined with a~dynamical meteoroid model of Jupiter-family comets (JFCs) that covers four orders of magnitude in particle diameter to investigate the consequences of grain-grain collisions in the inner Zodiacal Cloud. 
We compare this model to a~suite of observational constraints from meteor radars, the Infrared Astronomical Satellite (IRAS), mass fluxes at Earth, and inner solar probes, and use it to derive the population and collisional strength parameters for the JFC dust cloud. We derive a critical specific energy of $Q^*_D=5\times10^5 \pm 4\times10^5~R_\mathrm{met}^{-0.24}$~J~kg$^{-1}$ for particles from
Jupiter-family comet particles, making them 2-3 orders of magnitude more resistant to collisions than previously assumed. We find that the differential power law size index $-4.2\pm0.1$ for particles generated by JFCs provides a good match to observed data. Our model provides a good match to the mass production rates derived from the Parker Solar Probe observations and their scaling with the heliocentric distance. 
The higher resistance to collisions of dust particles might have strong implications to models of collisions in solar and exo-solar dust clouds. The migration via Poynting-Roberson drag might be more important for denser clouds, the mass production rates of astrophysical debris disks might be overestimated, and the mass of the source populations might be underestimated. Our models and code are freely available online.
\end{abstract}


\section{Introduction} \label{SEC:INTRO}
Despite its tenuous appearance (we can see stars and galaxies from our planet), solar system dust reveals its presence via collisions. Many spacecraft have been directly affected or damaged by particle impacts \citep{Drolshagen_Moorhead_2019, Szalay_etal_2020, Malaspina_etal_2022, Williams_etal_2017} and consequential plasma generation \citep{Close_etal_2010}. Surfaces of airless bodies are continuously eroded, a process thought to produce  tenuous extraterrestrial exospheres \citep{Killen_Hahn_2015, Pokorny_etal_2019, Pokorny_etal_2021}. 

The mutual collisions of dust and meteoroids have been studied for decades, sometimes controversially.
Some studies predict that the collisions are a main driver of dust/meteoroid removal in the inner solar system \citep[e.g.,][]{Grun_etal_1985,Mann_etal_2004}, whereas other studies show that collisions cannot be as frequent as previously thought if models are to match the shape and distribution of the Zodiacal Cloud \citep[e.g.,][]{Nesvorny_etal_2011JFC, Pokorny_etal_2014, Soja_etal_2019}. 

Before the advent of high-performance computing, mutual meteoroid collisions were examined analytically. 
Such analytic treatments have succeeded in reproducing several observed meteoroid-related phenomena. For instance, \citet{Wiegert_etal_2009} implemented a hybrid of \citet{Grun_etal_1985} and \citet{Steel_Elford_1986} collisional lifetimes to support a dynamical model that was able to model certain aspects of the orbital distribution of radar meteors at Earth using cometary stream modeling.

Simple analytic treatments produced some surprises as well.
\citet{Nesvorny_etal_2011JFC} showed that implementing the nominal collisional lifetime of \citet{Grun_etal_1985} in their dynamical model resulted in the removal of the majority of radar-detectable meteoroids before they reach Earth; the authors estimated that collisional lifetimes $30-100\times$ longer are necessary to reproduce the observed orbital distributions of meteors at Earth. \citet{Pokorny_etal_2014} implemented the \citet{Steel_Elford_1986} collisional model to the population of meteoroids originating from Halley-type comets, which is characterized by a broad range of orbital eccentricities. \citet{Pokorny_etal_2014} found a result like that of \citet{Nesvorny_etal_2011JFC}: reproducing these long-period-comet meteoroids required collisional lifetimes $20-50\times$ longer than those estimated in \citet{Steel_Elford_1986}. 
\citet{Soja_etal_2019}, in generating the latest version of the Interplanetary Meteoroid Environment Model, concluded that two different collisional lifetime categories provide the best fit to the Zodiacal Cloud measurements; for meteoroids with diameters $D<125~\mu$m, the \citet{Grun_etal_1985} collisional lifetimes were found to be adequate, while for $D>125~\mu$m, the authors required collisional lifetimes that are 50 times longer. Because smaller particles have shorter dynamical lifetimes \citep[due to the greater importance of Poynting-Robertson drag at small sizes;][]{Burns_etal_1979}, they might be insensitive to collisions even for the nominal collisional lifetimes estimated from \citet{Grun_etal_1985}.

The two most commonly used methods for estimating collisional lifetimes were developed in \citet{Grun_etal_1985} and \citet{Steel_Elford_1986}. Both of these works assume a certain shape, density, and velocity distribution of particles in the Zodiacal Cloud  based on observations available at the time, and calculate the collisional lifetime of a meteoroid on a given orbit. There are two major differences between these two approaches. First, \citet{Grun_etal_1985} characterized meteoroid strengths by a binding energy that the kinetic energy of a projectile must exceed in order to cause catastrophic destruction, whereas \citet{Steel_Elford_1986} only required a target-to-projectile radius ratio of $\sim30-40$. Second, the \citet{Steel_Elford_1986} collisional lifetimes are dependent on a meteoroid's orbital inclination, while in \citet{Grun_etal_1985} the collisional lifetimes are only a function of a meteoroid's heliocentric distance and mass.

Each approach has flaws. Disregarding the way in which impactor energy scales with heliocentric distance, as \cite{Steel_Elford_1986} do, can over/underestimate the rates of destructive collisions in the closest/farthest regions of the Zodiacal Cloud \citep{Mann_etal_2004}. Neglecting inclination in the manner of \cite{Grun_etal_1985} can result in underestimating the collisional lifetimes of meteoroids on highly-inclined orbits and overestimating the lifetimes of retrograde particles with orbits close to the ecliptic \citep{Steel_Elford_1986}. Note that \citet{Grun_etal_1985} and \citet{Steel_Elford_1986} provide similar estimates for collisional lifetimes for meteoroids with orbits close to the ecliptic at 1~au. Finally, any approach to collision modelling that is not fully three-dimensional underestimates the effects of mean motion resonances \citep{Nesvold_Kuchner_2015}  and secular forcing \citep{Nesvold_Kuchner_2015_SMACK} on collision rates.

To overcome the various drawbacks and assumptions of analytic methods to estimate the effects of mutual dust and meteoroid collisions, \citet{Stark_Kuchner_2009} devised a numerical algorithm for treating collisional evolution of any particle cloud. This new ``collisional grooming" algorithm is a self-consistent method that uses a three-dimensional distribution of model particles and particle trajectories.
%Understanding various setbacks of analytic treatment of collisions \citet{Stark_Kuchner_2009} inspired by \citet{Krivov_etal_2005} collisional model devised an algorithm for 3D collisional evolution of a particle cloud. 
The \citet{Stark_Kuchner_2009} algorithm does not make any assumptions about the shape of the dust cloud or the collisional lifetime but rather uses the modeled dust spatial and velocity distribution to estimate the rate of collisions at each point in the cloud. Using a collision-less dust cloud model as a starting point, the method iteratively applies a collision correction to the dust cloud density until a convergent steady-state solution is achieved: that is, the collision-less dust cloud groomed by the steady-state solution yields the same steady-state solution. The \citet{Stark_Kuchner_2009} method has been applied to model dust in the outer solar system \citep{Kuchner_Stark_2010, Poppe_2016} where the quantity of model constraints is scarce. This method has not previously been applied to the inner, more easily observable, regions of the Zodiacal Cloud.

Motivated by the surprisingly large collision lifetimes required by previous models and by the abundance of different dust and meteoroid related constraints available for the inner solar system, we set ourselves the following goals: (1) create a publicly available version of the \citet{Stark_Kuchner_2009} code that can handle large dynamical dust models able to reproduce various inner solar system phenomena, (2) find the best fit to currently available constraints in the inner solar system using a new model of Jupiter-family comet (JFC) meteoroids, (3) compare the collisional lifetimes estimated from our new model to those used in previous studies, and (4) constrain the material strength of JFC meteoroids by using the collision rates of our best-fitting model to constrain their binding energy.

\section{Meteoroid material characteristics and collision conditions}
\label{SEC:Collision_Conditions}
When a meteoroid (target) collides with a member of the particle cloud (projectile), this event has three possible outcomes: (1) the target \replaced{is unchanged}{suffers negligible damage}; (2) the target incurs partial damage; and (3) the target is catastrophically destroyed leaving behind a cloud of daughter particles. In this article we only consider outcome (1) and (3), where we assume that the cloud of daughter particles is blown out of the solar system due to radiation pressure and thus can be neglected. We neglect the second scenario, the accumulation of damage on each particle, due to unknown effects of particle-particle collisions on the immediate structure of individual particles  \citep[as suggested in][]{Grun_etal_1985}. 

In our analysis, we only check for catastrophic disruption as though the particle is a fresh particle, regardless of its dynamical age. The particle erosion effect would in general decrease the particles' collisional lifetimes, but due to the lack of any empirical/laboratory constraints we reserve that topic for future work.

Our catastrophic collision conditions follow the formalism from \citet{Krivov_etal_2005}, in which a catastrophic collision occurs when the kinetic energy of the impactor $E_\mathrm{imp}$ is greater than or equal to the target's binding energy $E_\mathrm{bind}$:
\begin{eqnarray}
    E_\mathrm{imp} & = & \frac{1}{2}m_\mathrm{imp} v_\mathrm{coll}^2, \\
    E_\mathrm{bind} & = & A_s \left(\frac{R_\mathrm{tar}}{1\mathrm{m}} \right)^{B_s} m_\mathrm{tar} = Q^*_D m_\mathrm{tar} , 
    \label{EQ:Binding_Energy}
\end{eqnarray}
where $A_s$ and $B_s$ are material constants, $m_\mathrm{tar}$ is the mass of the target, $R_\mathrm{tar}$ is the radius of the target, $m_\mathrm{imp}$ is the mass of the impactor, \deleted{and} $v_\mathrm{coll}$ is the relative collision velocity\added{, and $Q^*_D$ is the critical specific energy}. In this article, we \deleted{exclusively} use MKS units unless explicitly stated otherwise (e.g., for the semimajor axis we use astronomical units).

\citet{Krivov_etal_2006} set $A_s$ so that $Q^*_D = A_s \left(R_\mathrm{tar}/1\mathrm{cm}\right)^{B_s} = 10^6$ erg g$^{-1}$ for a 1 meter target and set $B_s = -0.24$.
These choices correspond to $A_s = 3.02\times10^6$ erg g$^{-1}$ in CGS units or $A_s = 100.0$ J kg$^{-1}$ in MKS units. For comparison, \citet{Grun_etal_1985} uses the following expression for their binding energy:
\begin{align}
    % E_{\mathrm{bind}, \, \mathrm{G85}} &= 4.6\times 10^{6} S_c^{0.45}R_\mathrm{tar}^{-0.225} m_\mathrm{tar} ~[\mathrm{CGS}] = 163.2S_c^{0.45}R_\mathrm{tar}^{-0.225} m_\mathrm{tar} ~[\mathrm{MKS}],
    E_{\mathrm{bind}, \, \mathrm{G85}} &= 4.6 \times 10^{6} ~\mathrm{cm}^2~\mathrm{s}^{-2} \cdot {S_c}^{0.45} 
    \left( \frac{m_\mathrm{tar}}{\textrm{1 g }} \right)
    \left( \frac{R_\mathrm{tar}}{\textrm{1 cm}} \right)^{-0.225} ~[\mathrm{CGS}]
       \\
       & = 163.2~\mathrm{m}^2~\mathrm{s}^{-2} \cdot {S_c}^{0.45} 
    \left( \frac{m_\mathrm{tar}}{\textrm{1 kg}} \right)
    \left( \frac{R_\mathrm{tar}}{\textrm{1 m }} \right)^{-0.225} ~[\mathrm{MKS}]
           \label{EQ:Binding_Energy_Grun}
\end{align}
where $S_c$ is the unconfined compressive strength in kilobars (kbar), which is typically between 0.5-4 kbar or 50-400 MPa \citep{Ostrowski_Bryson_2019}.
Comparing this expression to Eq.~\ref{EQ:Binding_Energy} assuming $S_C^{0.45} \approx 2$, we see that \citet{Grun_etal_1985} used $A_s \approx 300$ J kg$^{-1}$, a value approximately three times larger than the \citet{Krivov_etal_2006} value ($A_s = 100$ J kg$^{-1}$). \added{\citet{Rigley_Wyatt_2021} in their model for model for JFC meteoroid and dust cloud find values of $A_s = 660$ and $B_s = -0.90$. Their value of $A_s$ is comparable to those of \citet{Grun_etal_1985}, however, their slope $B_s$ is much steeper as shown in comparison to other relevant works in their Figure 25}.

\begin{figure}
\epsscale{1.0}
\plotone{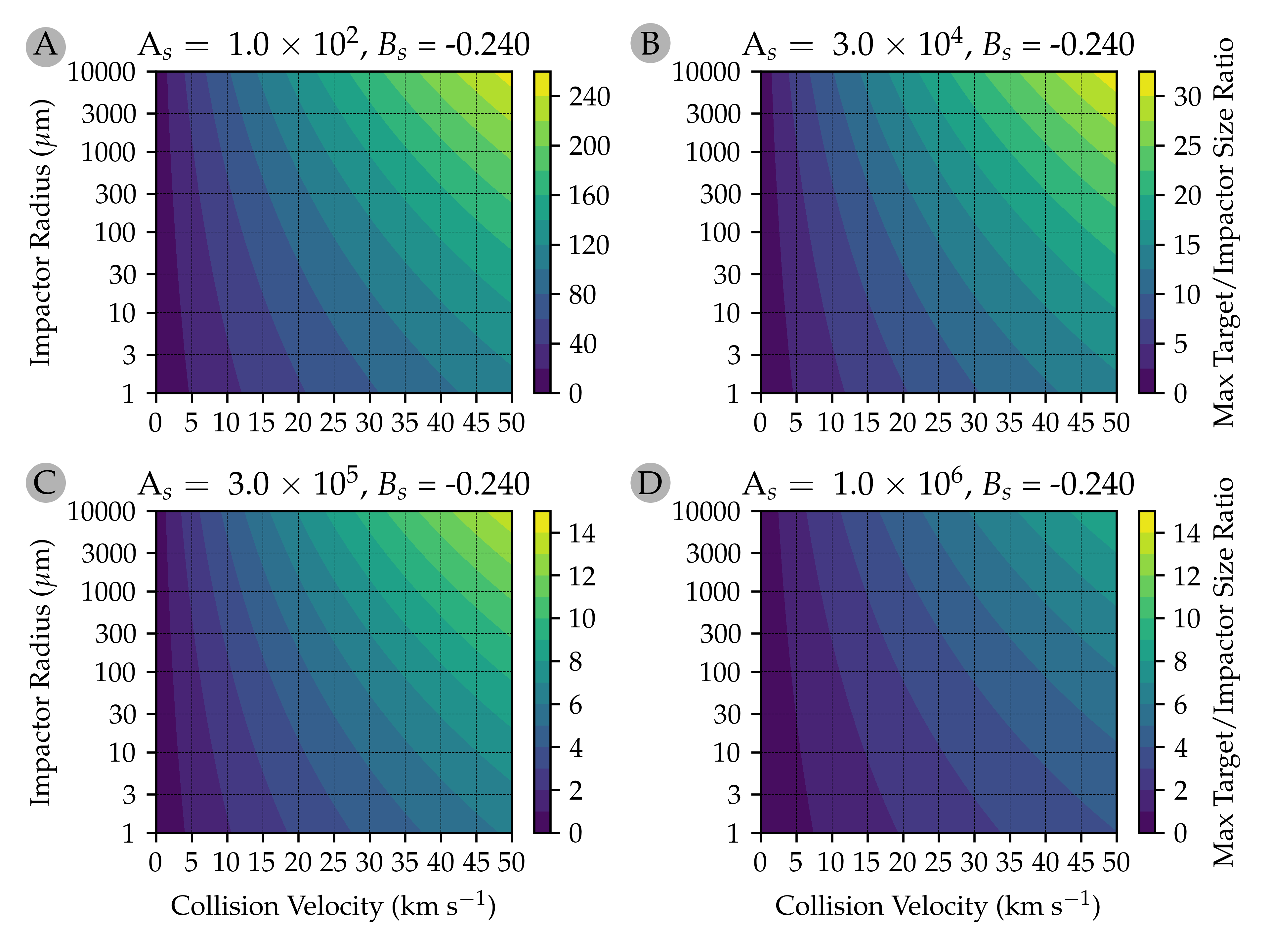}
\caption{\label{FIG:Collisions_Analytic}
Maximum target and impactor size ratio for destructive collisions $\mathcal{R}_\mathrm{max}$. If the target is larger than $\mathcal{R}_\mathrm{max}$ then the target is not destroyed during the collision. This Figure shows values of $\mathcal{R}_\mathrm{max}$ for four different values of $A_s$. In Panel A we show $A_s = 10^{2}$ J kg$^{-1}$, which is equal to $A_s = 3.02\times10^6$ erg g$^{-1}$ used in \citet{Krivov_etal_2005}. Panels B-D show $\mathcal{R}_\mathrm{max}$ for $A_s = 3\times10^4, 3\times10^5, \mathrm{~and~} 10^6$ J kg$^{-1}$ which are cases of interest in this manuscript.
}
\end{figure}

After reorganizing a few terms in Eq.~\ref{EQ:Binding_Energy}, we can obtain a ratio $\mathcal{R}_{\mathrm{max}}$ between the maximum destructable target radius $R_\mathrm{max,tar}$  for a given impactor radius $R_\mathrm{imp}$ and $v_\mathrm{coll}$:
\begin{equation}
   \mathcal{R}_\mathrm{max} = \frac{R_\mathrm{max,tar}}{R_\mathrm{imp}} = \left[\frac{R_\mathrm{imp}^{-B_s} v_\mathrm{coll}^2 }{2 A_s} \right] ^\frac{1}{3+B_s}.
   \label{EQ:Max_Target_Radius}
\end{equation}
Figure \ref{FIG:Collisions_Analytic} plots Eq.~\ref{EQ:Max_Target_Radius} for the range of impactor diameters modeled here and for collision velocities up to $v_\mathrm{coll} = 50$ km s$^{-1}$.
%%%%
%%%

\section{Collisional Grooming Code Description}

Motivated by the findings of \citet{Stark_Kuchner_2009} and \citet{Kuchner_Stark_2010}, we decided to implement a new version of the \citet{Stark_Kuchner_2009} collisional grooming algorithm that includes several optimizations and improvements. The entire code is written in C++ and is readily accessible at the project GitHub page%
%\footnote{\url{ https://github.com/McFly007/AstroWorks/}}
\footnote{\url{ https://github.com/AnonymizedForPeerReview}}
together with a simple test case particle model described in this section. The code does not currently support multi-CPU/GPU processing.\deleted{; however, this feature is being developed as part of our future work.}  The basic collisional algorithm is described in \citet{Stark_Kuchner_2009}, but we review it briefly here. 

\subsection{The Seed Model}

As an input, the grooming algorithm requires records of the dynamical evolution of a particle cloud in Cartesian coordinates: for each particle, the code requires a list of the particle's position and velocity vectors as it evolves over its entire lifetime, from creation to destruction, assuming no collisions. Such a list is commonly obtained by simulating a population of particles in a numerical $N$-body integrator as we describe in Sections \ref{SEC:SIMPLE_TEST} and \ref{SEC:MODEL_JFC}.

Each particle in an $N$-body simulation starts with some initial $\vec{r}$, $\vec{v}$ and then dynamically evolves through a complex pathway until it meets its end in one of several possible particle sinks (e.g., gets too far from the central star, disintegrates when too close to the central star, hits a planet/moon, etc.). Our grooming code has no requirements for the end conditions; its only requirement is that the particle properties are recorded at uniform time intervals $T_\mathrm{record}$. Likewise, the code does not \emph{a priori} expect that the particles' positions and velocities are recorded in either barycentric or asterocentric (star centered) coordinates, and can handle multi-star systems without any adjustments. This suite of particle records form the collision-less ``seed'' model.

Since the collisional grooming algorithm seeks a steady-state solution, the time domain and space domain merge. Each particle in the $N$-body simulation is replaced by a pathline representing many particles. Each of these pathlines is assigned a birth number density, $n_\mathrm{init}$, which decreases along its path due to collisions with other particles. 

%%% OLD VERSION
%Our code takes as input the orbital histories of particles recorded in Cartesian coordinates: the distribution of particle position and velocity vectors over time are traced through the cloud. %The time domain is important for the collisional grooming calculation, since each ``particle'' is assigned an initial number density, $n_\mathrm{init}$, that then continuously decreases over time due to collisions with other particles. 
All the particle records are distributed into a three-dimensional grid of bins of equal dimensions, creating a three-dimensional histogram that represents the total particle density of the cloud. Each bin can be imagined as a box that contains recorded information of all particles that happen to be inside the box. Our code stores the particle's velocity vector $\vec{v}$, diameter $D$, and number density $n$ in each bin, for all the particle pathlines that have records in that bin. 

%The time domain is not stored since we assume that the system is in a steady-state. The position of each particle is also not stored, since we know their position in the 3D grid and we assume that particles are uniformly distributed in space within the 3D bin. 

\subsection{Grooming With Fewer Iterations}
Once a collision-less seed model is loaded into the 3D histogram, we can investigate how collisions attenuate individual causal pathlines. When a particle with index $i$ passes through a particular 3D histogram bin located at $\vec{r} = (x,y,z)$ with the collisional depth $\tau_\mathrm{coll}(\vec{r})$, its pre-collision number density $n_\mathrm{pre,i}$ is attenuated as
\begin{equation}
   n_\mathrm{post,i} = n_\mathrm{pre,i} e^{-\tau_\mathrm{coll}(\vec{r})},
\label{eq:grooming}
\end{equation}
where $n_\mathrm{post,i}$ is the post-collision number density that particle with index $i$ carries to the next record interval. We refer to the process of adjusting the number densities via Equation~\ref{eq:grooming} as ``grooming''. \added{The value of $\tau_\mathrm{coll}$ effectively describes the number of collisions a particle experiences per sampling interval and depends on the model setup.}

The collisional depth is calculated in each bin in our code as
\begin{equation}
    \tau_\mathrm{coll}(\vec{r}) = \sum_{k} n_k(\vec{r}) \sigma_{i,k} v_{\mathrm{rel}_{i,k}}(\vec{r}) T_\mathrm{record},
\end{equation}
where $n_k$ is the particle density of the $k$\textsuperscript{th} particle record in the bin, $\sigma_{i,k} = \pi(R_i+R_k)^2$ is the collisional cross section assuming that both particles have negligible escape velocities compared to their relative encounter velocities (no gravitational focusing), $R_i$ and $R_k$ are the radii of the investigated and $k$\textsuperscript{th} particle respectively, $v_{\mathrm{rel}_{i,k}}$ is the relative velocity between the investigated and $k$\textsuperscript{th} particle, where the sum is over all $k$ records within an investigated bin. The collisional lifetime $T_\mathrm{coll}$ -- i.e., the collision e-folding lifetime -- is defined as
\begin{equation}
    T_\mathrm{coll} = \frac{T_\mathrm{record}}{\tau_\mathrm{coll}},
    \label{EQ:Collisional_Lifetime}
\end{equation}
and describes the timescale on which the weight of the particle cloud is decreased to $1/e=36.79\%$ of the original weight; that is, $63.21\%$ of particles are destroyed after one $T_\mathrm{coll}$. These approximations only work if the record sampling frequency is much greater than the average time between collisions, and if the bins are small enough to resolve any 3-D structures in the cloud, including those created by collisions. If the bins are too large, the code can over-estimate the rate of collisions and skew the final results.

All particle pathlines are processed independently: each one of them is individually groomed using the same initially collision-less particle cloud, which results in the first iteration of a collisionally processed particle cloud. However, since all particles are attenuated based on the collision-less model, this first iteration overestimates the collisional depths in each bin and arrives at a solution that underestimates the particle densities in each bin. Therefore, we perform further iterations of the collisional grooming process until the pre- and post-iteration state difference is smaller than a set precision. \citet{Stark_Kuchner_2009} showed that such an iterative process jumps between overestimated and underestimated states but can converge to a steady-state solution. 

However, we find that this convergence can be slow or non-existent for dense particle clouds with high values of collisional depth $\tau_\mathrm{coll}$. Therefore, we add an additional step to the process beyond the  \citet{Stark_Kuchner_2009} algorithm.  After each iteration, we calculate the difference measure $\mathcal{D}$ between the pre- and post-iteration number density for each particle in the particle cloud.  Since we know the number density $n$ of each particle before and after the collisional iteration step, we get
\begin{equation}
\mathcal{D} = \max{\left|\log_{10}\left(\frac{n_{\mathrm{pre}_{k}}}{n_{\mathrm{post}_{k}}}\right)\right|},    
\end{equation}
where $k$ iterates over all particle records in our 3D grid. To decrease the effect of over/underestimating the collisional depths of the cloud, we combine the pre- and post-iteration states of the particle clouds and assigning all particles in the post-iteration state the following number density:

\begin{equation}
    n_{\mathrm{new}_{k}} = \mathcal{S} n_{\mathrm{post}_{k}}+(1-\mathcal{S})n_{\mathrm{pre}_{k}},
\end{equation}
where
\begin{equation}
\mathcal{S} = \left\{\begin{array}{ll}
0.3 (1 - \tanh{(0.1\mathcal{D})}) & \mathcal{D}\ge 1\\
0.3 & \mathcal{D} < 1
\end{array}\right.
\label{EQ:Weighting_Form}
\end{equation}
Equation \ref{EQ:Weighting_Form} expresses what weight the pre- and post-iteration state hold. We arrived at this prescription by testing various functional forms of $\mathcal(S)$ and various values of $\mathcal{D}$. These different forms involved linear combinations of linear, quadratic, cubic, exponential, logarithmic, trigonometric, and hyperbolic functions. We find that the hyperbolic tangent used in Eq. \ref{EQ:Weighting_Form} works best in various tests conducted in Sec. \ref{SEC:SIMPLE_TEST}. Dust clouds with different particle distribution or densities could benefit from more customized re-weighting factors. If the final state of the collisionally groomed cloud is approximately known, then it can be prescribed in the first iteration to speed-up the computation significantly. 

Compared to the original \citet{Stark_Kuchner_2009} setup, our approach converges 10-100 times quicker for thick clouds with frequent collisions ($\tau_\mathrm{coll}>10^{-3}$), when the difference for the cloud state before and after iteration is high, i.e. for $\mathcal{D}>0$.  The algorithm also performs better than \citet{Stark_Kuchner_2009} even for particle clouds with fewer collisions; however, the improvement is less significant in these cases because they only tend to require only several ($<10$) iterations.

\section{Simple test case}
\label{SEC:SIMPLE_TEST}
We tested our code on a small dataset, reproducing a test from Section 2.3.5 of \citet{Stark_Kuchner_2009}. We simulated the evolution of a cloud of 10,000 particles orbiting a Sun-like star with no planets, where all particles are released in the form of a thin circular ring. For this particular seed model, the effects of collisions can be expressed analytically, as shown in \citet{Wyatt_1999PHD}. All particles in our model were $D=240~\mu$m in diameter and had bulk densities of $\rho = 2000$~kg~m$^{-3}$; that is, the ratio of radiation pressure to gravity was $\beta =2.375 \times 10^{-3}$. 

We express the effects of radiation pressure and Poynting-Robertson drag on dust particles in terms of the parameter $\beta$, which is the ratio between the radiation pressure force $F_\mathrm{rad}$ and the gravitational force of the central star $F_\mathrm{grav}$:
\begin{equation}
    \beta = \frac{F_\mathrm{rad}}{F_\mathrm{grav}}
    = \frac{1.14\times10^{-3}Q_\mathrm{pr}}{\rho D} \frac{L_\star}{L_\odot}  \frac{M_\odot}{M_\star}
    = 1.14\times10^{-3}Q_\mathrm{pr}{\left(\frac{1~\mathrm{kg}~\mathrm{m}^{-3} }{\rho}\right) \left(\frac{1~\mathrm{m}}{D}\right)}\frac{L_\star}{L_\odot}  \frac{M_\odot}{M_\star},
    \label{EQ:Beta}
\end{equation}
where $L_\star$ is the central star luminosity, $L_\odot = 3.828 \times 10^{26}$ W is the solar luminosity, $M_\star$  is the mass of the central star, and $M_\odot = 1.98847 \times 10^{30}$ kg is the solar mass, \textbf{$Q_\mathrm{pr}$ is the radiation pressure coefficient that we assume to be unity ($Q_\mathrm{pr}=1$)}. For a detailed description of the effects of stellar radiative forces on the dynamics of dust particles and meteoroids see \citet{Burns_etal_1979} or \citet{Murray_Dermott_1999_Book}.
For the rest of this paper, we assume a density of $\rho = \textrm{2,000~kg~m}^{-3}$, the value used in \citet{Nesvorny_etal_2011JFC}. Note that particle density influences both the particle-particle collision criteria as well as constrainable quantities such as the mass flux, particle cross-section, etc.

Using the same initial conditions as in \citet{Stark_Kuchner_2009}, all particles in our test case model were released with an initial semimajor axis of $a = 10$~au, on circular orbits with eccentricity $e=0$, and inclinations $I = 10^\circ$. Their initial longitudes of the ascending node $\Omega$, arguments of pericenter $\omega$, and mean anomaly $M$ were picked randomly between $0$ and $360^\circ$. On their release, all particles are pushed to slightly more eccentric orbits due to radiation pressure. Then they spiral toward the central star via PR drag \citep{Burns_etal_1979}. We recorded the particle position and velocity vectors every 6956 years until the particles reached a heliocentric distance of 0.5~au. We used the \texttt{SWIFT\_RMVS\_3} numerical integrator \citep{Levison_Duncan_2013} to create the seed model. For the collisional grooming, we selected a bin size of 0.05~au where bins extend up to 10 au from the central star. 

\begin{figure}
\epsscale{1.0}
\plotone{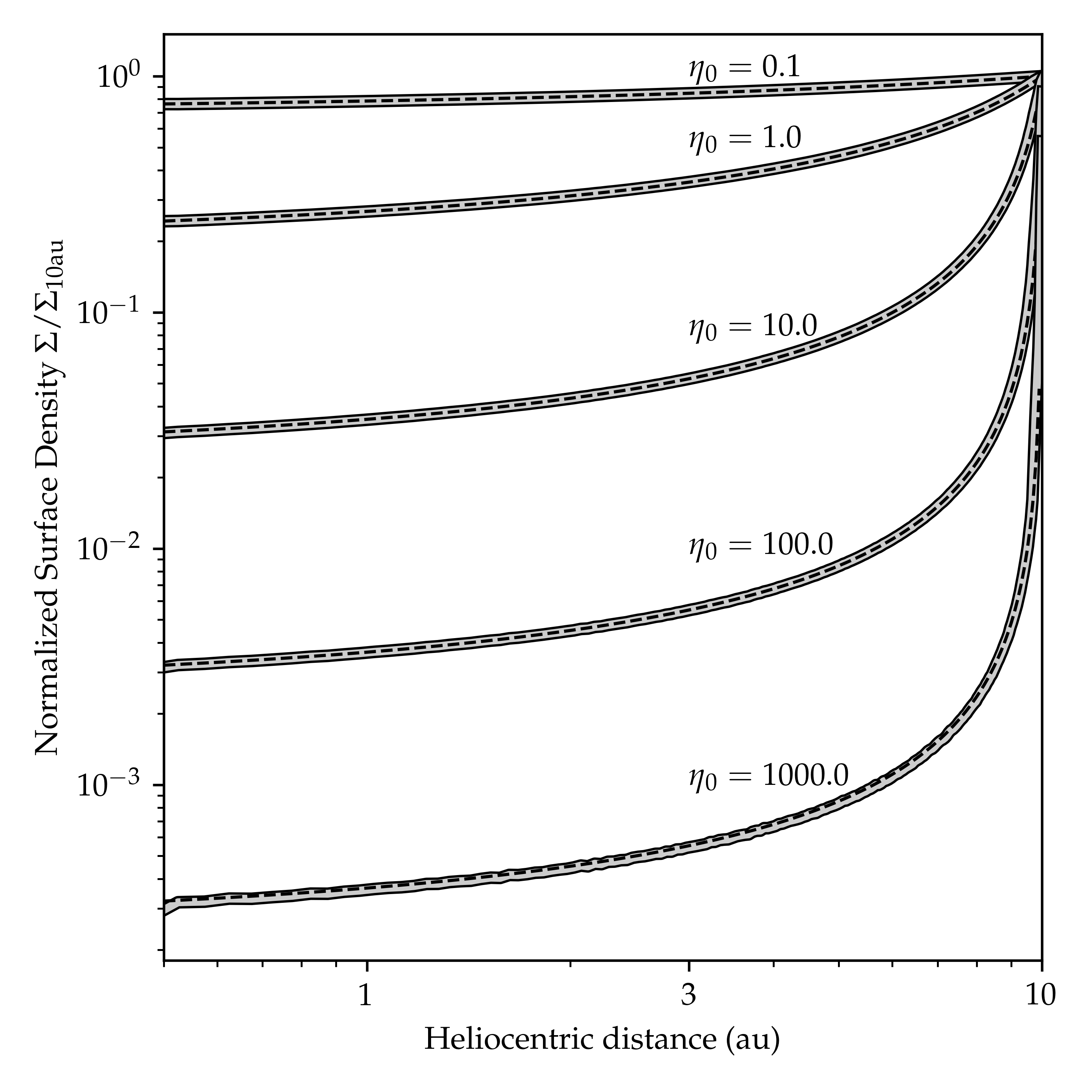}
\caption{\label{FIG:Simple_Stark_Test}
Normalized surface density $\Sigma/\Sigma_{10\mathrm{au}}$ of the steady-state solution as a function of heliocentric distance for our simple test scenario of dust migrating from heliocentric distance of 10 au via PR drag with no planets included (gray solid line). All collisions between particles are destructive in this case and the collision products are omitted. The analytic solution given by Eq.~\ref{EQ:Simple_Analytic} is represented by a black dashed line. We ran our collisional grooming code for five different values of $\eta_{0}$, which is linearly proportional to the collision-less cloud mass, surface area, or the total number of particles. The analytic solution agrees with the solutions of our code for all five values shown here.
}
\end{figure}

Figure \ref{FIG:Simple_Stark_Test} shows the final distribution of normalized surface density $\Sigma/\Sigma_\mathrm{10au}$ as a function of heliocentric distance $r_\mathrm{hel}$ produced by our new, improved collisional grooming algorithm. Our simple scenario closely resembles the one-dimensional analytic model in Eq. 3.36 of \citet{Wyatt_1999PHD}, which has a surface density of :
\begin{equation}
    \Sigma/\Sigma_\mathrm{10au} = \frac{1}{1+4\eta_0\left(1-\sqrt{r_\mathrm{hel}/r_0}\right)},
    \label{EQ:Simple_Analytic}
\end{equation}
where $\eta_0$ is the ratio between the PR decay time $t_\mathrm{PR}$ and the collisional lifetime $t_\mathrm{coll}$ at $r_0$. For $\eta_0 <1$, the collisional lifetime is longer than the PR drag timescale and the majority of particles spiral inward to the central star without being collisionally destroyed. On the other hand, for $\eta_0 \gg 1$, collisions dominate the dynamical evolution of particles and only a small fraction of the original population survives long enough to reach small heliocentric distances. 

Figure \ref{FIG:Simple_Stark_Test} compares the analytic prediction from Eq.~\ref{EQ:Simple_Analytic} (dashed black line) to the results produced by our collisional grooming code for the test case (gray solid line) for five different values of $\eta_0$ spanning 5 orders of magnitude. Figure \ref{FIG:Simple_Stark_Test} shows that our code correctly estimates the rates of collisions in this simple scenario and matches predicted profiles.

For $\eta_0 = 1000.0$, the particle cloud is very dense and the rate of collisions is approaching the resolution limit; you can see slight differences between the analytic model and the numerical model toward the center of the cloud. These errors can be alleviated by running the code on a finer grid with more particles or with higher recording frequency (shorter $T_\mathrm{record}$). Similar limitations were observed in \citet{Stark_Kuchner_2009}. 
Such high values of $\eta_0$ can be found in  massive debris disks such as the one orbiting $\beta$ Pictoris, or for centimeter sized particles. 
%that are observable with current instruments \citep{Wyatt_2005}. In these dust clouds the collisions can dominate the radiative driven particle dynamics and make the effects of PR drag negligible. 
However, in real-life scenarios the dust clouds are composed of particles with (A) a wide range of sizes and (B) different compositions, where both of these factors can change the value of $\eta_0$ by orders of magnitude even within the same dust population. 
%For these reasons any complex scenario cannot be represented by Eq.~\ref{EQ:Simple_Analytic} and must be treated as a whole considering all particle sizes and compositions at once. The only debris cloud with enough observations and reproducible constraints is our own Zodiacal Cloud, which will be our main focus in the rest of this article.

%%%%
%%%

\section{Collisional grooming of Jupiter-family Comet meteoroids}

The main goal of this article is to study how collisions affect the main component of the Zodiacal Cloud: the cloud of dust and meteoroids originating from Jupiter Family Comets (JFCs). Numerous independent models show that JFC meteoroids dominate the flux and number density of the Zodiacal Cloud over a wide range of particle sizes in the inner solar system \citep{Koschny_etal_2019}.
%Visual observations of the innermost region of the Zodiacal Cloud linked its low-inclination component to dust and meteoroids originating in main belt asteroids and JFCs \citep{Hahn_etal_2002}. 
Infra-red observations from IRAS and COBE were best reproduced when the major component of the Zodiacal Cloud beyond 1~au originates from JFCs \citep{Nesvorny_etal_2010,Nesvorny_etal_2011JFC}. Indirect observations of cosmic dust entering the Earth's exosphere point to a dominant JFC component as well \citep{CarrilloSanchez_etal_2016, CarrilloSanchez_etal_2020}. Multiple independent long-term observations of radar meteor orbit radars showed the dominance of a short-period cometary component in the sporadic meteoroid background observed at Earth \citep{CampbellBrown_2008,Galligan_Baggaley_2004,Janches_etal_2015}. 

In this section, we present a ``seed'' model of JFC meteoroids that is an extension of the \citet{Nesvorny_etal_2011JFC} model.  We add collisions to this new dynamical model using our newly developed collisional grooming code. We compare this model to various independent observations to find a unique, best fit to the data. This process allows us to characterize the population of JFC meteoroids and the material parameters that govern their collisions, $A_s$ and $B_s$ (see Eq.~\ref{EQ:Binding_Energy}).

\subsection{Model of Jupiter-family comet meteoroids}
\label{SEC:MODEL_JFC}
To cover the large size range needed to reproduce our various constraints and the effects of collisions between dust particles and meteoroids originating in JFCs, we need to construct a new dynamical model.
%, an extension of \citet{Nesvorny_etal_2011JFC}. 
The models presented in \citet{Nesvorny_etal_2011JFC, Pokorny_etal_2018, Pokorny_etal_2019, Soja_etal_2019} all suffer from several shortcomings: (1) they do not cover a wide range of particle diameters, where for collisional purposes we require particles as small as 1 micrometer and as large as 1 centimeter (10,000 micrometers); (2) they have disproportionate and sparse size bins; and (3) they select source particles to integrate that have been launched into bound orbits, which biases the model.

This pre-selection of particles on bound orbits is usually performed to obtain a better yield of numerical simulations with respect to the computational time. However, this improved yield comes at a cost.
 The blow-out heliocentric distance is $r_\star = 2\beta a$; thus, the particles removed from simulations by the preselection process tend to be smaller ones  \citep{Moorhead_2021}.
These small particles ($\beta-$meteoroids) can disrupt larger particles, so omitting them could severely distort the Size-Frequency Distribution (SFD). This practice might have caused earlier models to over-predict the number fluxes for micron-sized particles observed by various spacecraft \citep{Malaspina_etal_2020}.

\begin{figure}
\epsscale{0.93}
\plotone{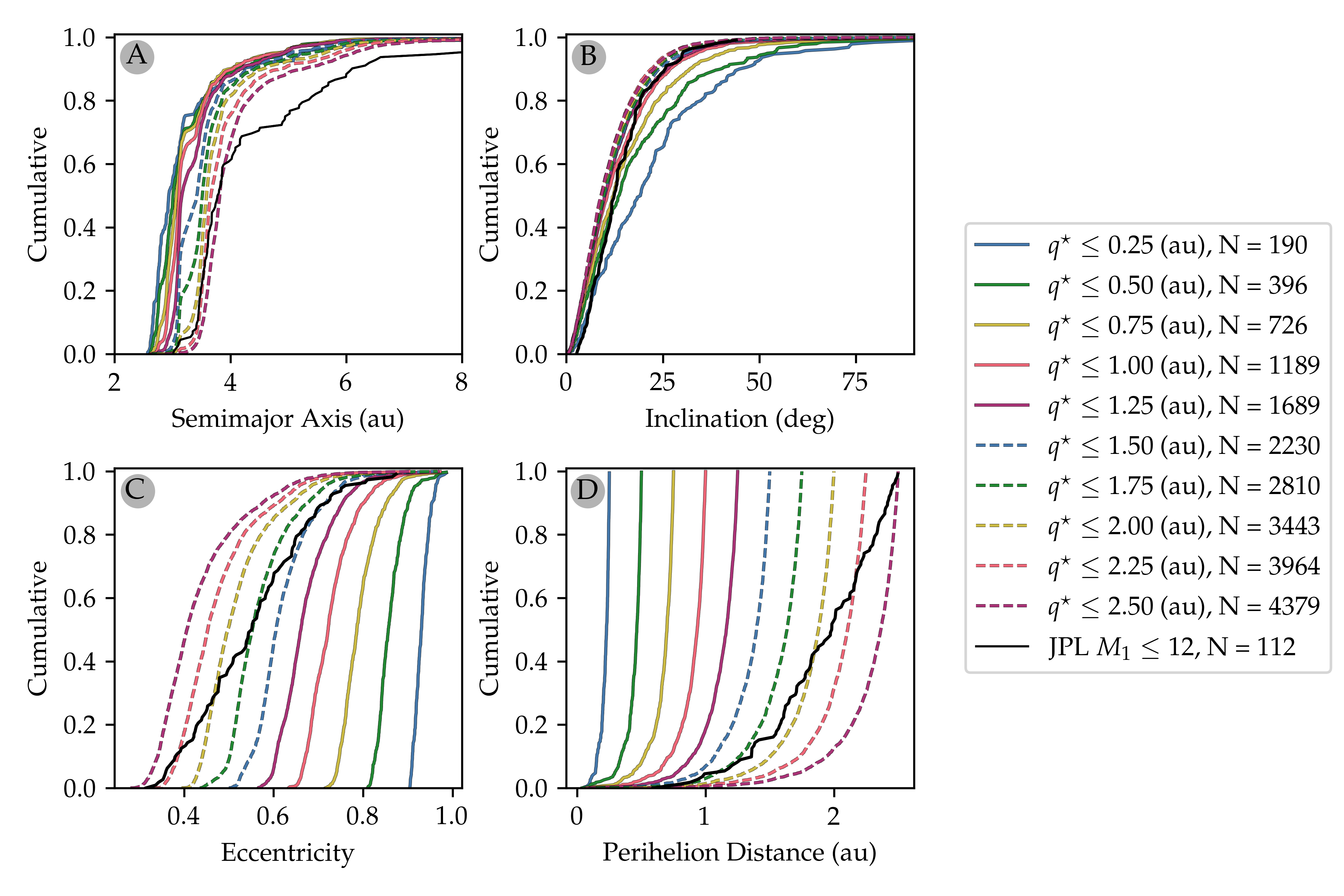}
\caption{
\label{FIG:JFC_SETUP}
Initial conditions for JFCs in our numerical models. Panel A shows the cumulative distributions of semimajor axis $a$ for 10 different values of critical perihelion distance $q^\star$ (color-coded solid and dashed lines) and the currently observed distribution of JFCs with comet total magnitude parameter $M_1 \le 12$. Panels B-D are the same as panel A but show instead the inclination (panel B), eccentricity (panel C), and perihelion distance (panel D) cumulative distributions. The legend also shows the total number of bodies for each perihelion distance set, $N$. The data for currently observed JFCs were obtained on May 20th, 2021 from \url{https://ssd.jpl.nasa.gov/sbdb_query.cgi}. ASCII version of the cumulative plots is available at the project GitHub.
%\footnote{\url{ https://github.com/AnonymizedForPeerReview}}.
%\footnote{\url{ https://github.com/McFly007/AstroWorks/}}
}
\end{figure}

To address shortcomings (1) and (2), we created a model that covers a large range of particle diameters $D$. We selected $\beta$ values for our modeled population of JFC dust and meteoroids ranging from $\beta=0.837$ ($D=0.6813~\mu$m) to $\beta = 5.7 \times 10^{-5}$ ($D=10,000~\mu$m) in 26 logarithmically spaced bins: 
\begin{equation}
   \beta =\frac{1.14}{\rho} {10}^{-i/6}, \, i\in[-1,24] \textrm{; i.e., for }\rho = \textrm{2,000~kg~m}^{-3}, \, D = {10}^{i/6}.
\end{equation}
%The dynamics of dust particles in our model depend only on $\beta$/ For the rest of this text we assume that $\rho = \textrm{2,000~kg~m}^{-3}$, the value used in \citet{Nesvorny_etal_2011JFC}, where we note that the particle density influences both the particle-particle collision criteria as well as the constrainable quantities such as the mass flux, particle cross-section, etc.

To address shortcoming (3) of previous models (pre-selection into bound orbits), we launch particles from their parent bodies assuming that the parent bodies are uniformly spread in mean anomaly $M$, regardless of whether we expect the particles to remain bound.

Our model uses the same parent body distribution as \citet{Nesvorny_etal_2011JFC}. Parent body orbits are assumed to be the same as those of objects originating in the Kuiper belt and transferred into the Jupiter-family comet orbits \citep{Levison_Duncan_1997}. The orbital elements are recorded/saved once they reach critical perihelion distance $q^\star$; i.e., their perihelion distance $q<q^\star$. We use 10 different values of $q^\star \in [0.25,0.50,0.75,1.00,1.25,1.50,1.75,2.00,2.25,2.50]$ au. Cumulative distributions of semimajor axis $a$, inclination $I$, eccentricity $e$, and perihelion distance $q$ of parent bodies in each $q^\star$ bin are shown in Fig.~\ref{FIG:JFC_SETUP}. 

Each meteoroid in our simulations is released from a randomly selected comet from the parent body population in the $q^\star$ bin. The meteoroids are released with $a$, $e$, $I$ corresponding to their parent bodies, while the remaining orbital elements $\Omega, \omega, M$ are selected randomly between 0 and $2\pi$.
%Meteoroids in our simulations are released from comets/parent bodies with $a$, $e$, $I$ the parent body population in the $q^\star$ bin, while the remaining orbital elements $\Omega, \omega, M$ are picked randomly between 0 and $2\pi$. 
All meteoroids are assumed to be test particles and are subjected to the gravitational force of the sun and eight planets, the effects of radiation pressure, and PR drag. The integration time step is 1 day (86400 seconds), where close encounters with planets are handled separately by the \texttt{SWIFT\_RMVS\_3} integrator. The particles are numerically integrated until they are removed from the simulation once they 1) are close to the sun $(r<0.005)$ au; \added{or} 2) impact one of the eight planets; \replaced{and}{or} 3) are too far from the sun $(r>10,000)$ au. Meteoroids in our model can be released on initially hyperbolic orbits due to radiation pressure
%by which we account for a more realistic meteoroid production scenario
which corresponds to a more realistic meteoroid production scenario where the comets produce dust close to the Sun. We do not assume any dependence of particle production rate based on cometary orbit such as cometary activity models described in \citet{Jones_1995} or \citet{Crifo_Rodionov_1997} since \citet{Nesvorny_etal_2010} showed that cometary activity itself is insufficient to supply the current Zodiacal Cloud; cometary fragmentation, not solar-driven out-gassing, generates most of the dust \citep{Rigley_Wyatt_2021}. 

In order to obtain the same number of particle records in each diameter range, we increased the number of simulated particles in our simulations until we recorded 10~GB for each particle size bin and 1~GB per perihelion distance bin. This amounts to 260~GB of data in binary format, which is approximately $8.2\times 10^{9}$ records. Particles of each size and originating from different $q^\star$ populations (Fig. \ref{FIG:JFC_SETUP}) have vastly different dynamical timescales. This means that the initial number of particles in different size and perihelion bins can differ by orders of magnitude, since small particles disperse quickly from the system while the largest particles in our sample can evolve on timescales that are 5 orders of magnitude longer. Table~\ref{TAB:Initial_JFC_Numbers} shows the total number of released particles per each size and perihelion distance cut-off bin.  \added{The amount of data provided by our JFC model is more than sufficient for the purpose of this work. We ran several tests and even 5\% of the total amount of orbital records were sufficient to provide the same results as the full data set.}
% \clearpage

\movetabledown=40mm
\begin{rotatetable}
\begin{deluxetable*}{rrrrrrrrrrr}

\tablecaption{\label{TAB:Initial_JFC_Numbers} Initial number of particles in each bin of our dynamical model of JFC meteoroids.}
\tablenum{1}
\tablehead{
\colhead{Diameter}& \multicolumn{10}{c}{Perihelion distance cut-off $q^\star$ (au)} \\
{($\mu$m)}  & {0.25} & {0.50} & {0.75} & 
{1.00} & {1.25} & {1.50} & {1.75} & {2.00} & {2.25} & {2.50}
} 
\startdata
0.6813 & 	 4,533,921 & 	7,313,546 & 	31,571,532 & 	31,571,532 & 	31,571,532 & 	31,571,532 & 	31,571,532 & 	31,571,532 & 	31,571,532 & 	31,571,532 \\ 	
1.000 & 	 3,868,972 & 	1,806,057 & 	1,316,070 & 	1,326,277 & 	2,084,598 & 	3,150,065 & 	3,792,783 & 	9,852,590 & 	25,538,430 & 	31,567,273 \\ 	
1.468 & 	 3,441,864 & 	1,393,944 & 	952,834 & 	851,050 & 	846,218 & 	831,987 & 	820,798 & 	1,209,052 & 	1,825,130 & 	2,643,050  \\ 	
2.154 & 	 2,719,543 & 	1,061,270 & 	746,095 & 	612,250 & 	559,452 & 	550,513 & 	586,155 & 	659,775 & 	724,325 & 	635,675  \\	
3.162 & 	 1,946,814 & 	718,967 & 	509,874 & 	436,720 & 	410,653 & 	386,167 & 	385,632 & 	413,728 & 	450,921 & 	481,286  \\
4.642 & 	 1,206,908 & 	517,818 & 	352,809 & 	291,147 & 	278,144 & 	275,560 & 	269,498 & 	260,384 & 	266,614 & 	297,619  \\
6.813 & 	 826,345 & 	383,037 & 	253,069 & 	193,678 & 	173,696 & 	179,585 & 	188,932 & 	198,395 & 	207,313 & 	220,574  \\
10.00 & 	 608,876 & 	287,040 & 	192,948 & 	147,366 & 	127,945 & 	130,249 & 	144,673 & 	156,889 & 	161,648 & 	164,247  \\
14.68 & 	 459,837 & 	218,634 & 	149,735 & 	117,891 & 	101,851 & 	101,492 & 	113,629 & 	127,745 & 	130,117 & 	123,996  \\
21.54 & 	 346,151 & 	168,091 & 	118,135 & 	94,551 & 	81,768 & 	80,015 & 	91,536 & 	109,117 & 	110,620 & 	99,335  \\
31.62 & 	 263,524 & 	133,293 & 	95,218 & 	75,346 & 	65,627 & 	63,907 & 	72,393 & 	91,350 & 	98,264 & 	83,713  \\
46.42 & 	 196,068 & 	105,388 & 	78,910 & 	62,534 & 	53,784 & 	52,818 & 	58,923 & 	77,132 & 	83,466 & 	70,398  \\	
68.13 & 	 147,297 & 	84,441 & 	66,856 & 	53,291 & 	45,884 & 	44,214 & 	49,262 & 	64,481 & 	75,039 & 	61,208  \\	
100.0 & 	 116,249 & 	68,828 & 	56,859 & 	48,604 & 	39,697 & 	36,093 & 	41,723 & 	53,741 & 	63,724 & 	51,011  \\	
146.8 & 	 89,167 & 	56,389 & 	47,329 & 	44,056 & 	33,815 & 	29,643 & 	33,184 & 	41,587 & 	55,565 & 	47,037  \\	
215.4 & 	 72,406 & 	47,503 & 	41,262 & 	38,626 & 	28,285 & 	26,230 & 	25,624 & 	34,958 & 	49,643 & 	36,400  \\	
316.2 & 	 57,915 & 	42,058 & 	35,719 & 	33,868 & 	28,261 & 	19,798 & 	20,849 & 	30,056 & 	44,135 & 	29,302  \\	
464.2 & 	 49,119 & 	39,278 & 	33,556 & 	33,358 & 	22,275 & 	17,306 & 	16,508 & 	25,261 & 	56,549 & 	35,690  \\	
681.3 & 	 38,954 & 	29,305 & 	28,101 & 	28,006 & 	21,075 & 	15,139 & 	12,266 & 	20,906 & 	40,569 & 	34,591  \\	
1000 & 	 38,903 & 	33,040 & 	26,306 & 	28,036 & 	23,630 & 	13,386 & 	11,612 & 	17,229 & 	79,307 & 	46,396  \\
1468 & 	 30,964 & 	24,719 & 	29,484 & 	23,106 & 	28,144 & 	14,025 & 	8,338 & 	15,175 & 	27,453 & 	31,040  \\	
2154 & 	 20,990 & 	25,983 & 	21,961 & 	35,412 & 	17,941 & 	7,809 & 	5,225 & 	7,590 & 	27,944 & 	20,220  \\	
3162 & 	 36,721 & 	26,404 & 	33,535 & 	14,525 & 	10,038 & 	7,401 & 	5,459 & 	23,706 & 	49,114 & 	27,001  \\	
4642 & 	 33,325 & 	24,077 & 	26,453 & 	28,119 & 	16,959 & 	4,114 & 	12,198 & 	10,182 & 	37,119 & 	11,084  \\	
6813 & 	 16,815 & 	21,517 & 	10,235 & 	9,853 & 	6,645 & 	5,469 & 	2,144 & 	10,709 & 	30,234 & 	14,610  \\
10000 & 	 20,389 & 	8,903 & 	23,125 & 	36,725 & 	32,375 & 	1,079 & 	10,875 & 	3,186 & 	15,592 & 	22,527 \\
\enddata
\end{deluxetable*}
\end{rotatetable}

\subsection{Observational Constraints and Model Parameters}
\label{SEC:Constriants}
Before we discuss the collisional grooming of our JFC dynamical model, let us introduce the observational constraints and the goodness-of-fit functions we will use to measure how well our model fits these constraints.
\added{
The next five sections introduce the model constraints:
\begin{enumerate}
    \item Heliocentric distance dust density profile (Sec. \ref{SEC:Constraint_1})
    \item Mass accretion rate at Earth (Sec. \ref{SEC:Constraint_2})
    \item Orbital distribution of meteors at Earth (Sec. \ref{SEC:Constraint_3})
    \item Total cross-section of the Zodiacal Cloud (Sec. \ref{SEC:Constraint_4})
    \item Size-frequency distribution at Earth (Sec. \ref{SEC:SFD})
\end{enumerate}
}

\subsubsection{Heliocentric distance density dust profile}
\label{SEC:Constraint_1}
The heliocentric distance ($r_\mathrm{hel}$) profile observed by Helios space probes \citep{Leinert_etal_1981} shows that the surface brightness of zodiacal light in the direction perpendicular to the eclipcic plane scales as $S\propto R^{-2.3\pm 0.1}$, which conversely means that the particle density scales with the heliocentric distance as $n\propto r_\mathrm{hel}^{-1.3\pm 0.1}$. These values hold for heliocentric distances from $r_\mathrm{hel}=0.3$ au to $r_\mathrm{hel}=1.0$ au and for $z-$distances close to the ecliptic: $z<0.05$ au. These values were confirmed by the Parker Solar Probe \citep{Howard_etal_2019}. In our model, we calculate the dust cloud density $n$ near the ecliptic by averaging the two bins adjacent to $z = 0$ for each heliocentric distance $r_\mathrm{hel}$ in the model. Then we fit the single-power law to the density profile and obtain the model power law index $\mathcal{M}_R$ and compare it to the value measured by Helios and PSP, $\mathcal{O}_R = -1.3\pm 0.1$.

\subsubsection{Mass accretion rate at Earth}
\label{SEC:Constraint_2}
The mass of meteoroids accreted onto Earth has been estimated using various ground-based and space-borne experiments. \cite{Love_Brownlee_1993} estimated the accretion rate from impact rates on the Long Duration Exposure Facility (LDEF) satellite to $\dot{M}_\Earth = 109,500 \pm 54,800$ kg day$^{-1}$. This value was later re-interpreted using the hydrocode iSALE in \citet{Cremonese_etal_2012} and the accretion rate was decreased to $\dot{M}_\Earth = 20,300 \pm 2,700$~kg day$^{-1}$ if the mass accretion is dominated by main belt asteroid meteoroids, and to $\dot{M}_\Earth = 11,500 \pm 1,400$~kg day$^{-1}$ if the mass accretion is dominated by cometary meteoroids. \citet{Moorhead_etal_2020} revisited the LDEF impact record and added the results from the Pegasus experiment that in 1965 measured the flux of dust particles on 194.5 m$^2$ of collecting area. \citet{Moorhead_etal_2020} were not able to reconcile both impact experiments, noting the significant uncertainties connected with both experiments and complexity of the correct interpretation of acquired data.

Other ways to constrain the meteoroid mass flux at Earth use metal abundances in Earth's mesosphere \citep{CarrilloSanchez_etal_2016, CarrilloSanchez_etal_2020} or accumulation of ablated and non-ablated meteoric material in Antarctic ice-layers \citep{Rojas_etal_2021}. Like space-borne estimates, these methods are model-dependent. For instance, \citet{CarrilloSanchez_etal_2016} showed that using different size-frequency distributions, reflecting observations from IRAS and Planck infra-red observations of the Zodiacal Cloud, provides ${\sim}50\%$ different estimates for the meteoroid mass flux at Earth. However, these approaches have the advantage that they can differentiate between different meteoroid populations impacting Earth.

We selected the mass flux of JFCs at Earth $\mathcal{M}_M = 19,600 \pm 7,500$ kg day$^{-1}$ from \citet{CarrilloSanchez_etal_2020} as our primary mass accretion constraint. This measurement provides the most recent estimate and agrees within uncertainty ranges with \citet{Rojas_etal_2021} who analyzed a large collection of melted and unmelted meteoroids from the Concordia site in Antarctica. Note that \citet{CarrilloSanchez_etal_2016} provides a different value for JFC \replaced{accretion rate}{mass flux at Earth}, $\mathcal{M}_{\dot{M}} = 34,600 \pm 13,400$ kg day$^{-1}$, which we also consider in our analysis as an alternative value.

\subsubsection{Orbital distribution of meteors at Earth}
\label{SEC:Constraint_3}
Two meteor orbit surveys have acquired more than 5,000,000 orbits: the Canadian Meteor Orbit Radar \citep[CMOR;][]{CampbellBrown_2008} and the Southern Argentine Agile Meteor Radar \citep[SAAMER;][]{Janches_etal_2015}. Additionally, several other facilities have acquired hundreds of thousands of meteor orbits, including the Advanced Meteor Orbit Radar\citep[AMOR;][]{Galligan_Baggaley_2004}, and the Middle and Upper Atmosphere Radar \citep[MU;][]{Kero_etal_2012}. 
\citet{Jones_etal_2001,Nesvorny_etal_2011JFC} showed that JFC meteoroids preferably feed the helion and anti-helion sources observed at Earth; i.e., the meteors appearing to come from sun-ward and anti-sun-ward directions. While other meteoroid populations are detected at Earth as well, they are prominently concentrated into the apex region \citep[OCCs;][]{Nesvorny_etal_2011OCC} and north/south toroidal region \citep[HTCs;][]{Pokorny_etal_2014}; see also \citet{Jones_2004_NASA}. The main belt asteroids release meteoroids that might appear in helion/anti-helion sources. However, the impact directions of these meteoroids are scattered over a wide range of angles around the celestial sphere and difficult to detect by radars due to their relatively small Earth impact velocities \citep{Wiegert_etal_2009}.  The main belt asteroids dominate the number flux of sporadic centimeter+ sized meteoroids ($D>10,000~\mu$m) at Earth \citep{Shober_etal_2021}. However, that size range is currently beyond the scope of our model; we can assume that the meteors observed by meteor radars in helion/anti-helion sources are dominantly sourced from JFCs. 

One of the significant differences between CMOR and SAAMER is in their transmitted power and the beam pattern, which translates into a difference in minimum detectable meteoroid mass. SAAMER, with its 60~kW transmitter and more focused beam, can see approximately an order of magnitude smaller particles in diameter than CMOR's 6/12 kW antennas with a wide beam pattern \citep{Weryk_Brown_2012}. Smaller particles suffer fewer collisions during their dynamical lifetimes than their larger counterparts \citep[see e.g.,][for different meteoroid sizes at Mercury]{Pokorny_etal_2018}. This effectively means that while SAAMER provides valuable data about meteoroid environment, it cannot constrain the effects of collisional grooming as efficiently as CMOR.  \citet{Nesvorny_etal_2011JFC} showed this contrast between AMOR and CMOR. AMOR with its narrow beam had a similar sensitivity to SAAMER and recorded meteors that showed almost no signs of collisional erosion. AMOR meteors were recorded with low eccentricity orbits resulting from efficient PR drag. On the other hand, CMOR meteors were mostly recorded having highly eccentric orbits. This shows that larger meteoroids detected by CMOR are collisionally removed before their orbits are circularized via PR drag and only the dynamically young population with high-$e$ orbits is detected at Earth.
%, showing that those meteoroids on circularized orbits suffered collisional evolution before arriving at Earth.  

For these reasons we use \citet{CampbellBrown_2008} helion and anti-helion raw orbital element distributions as our primary meteor orbit distribution constrains. These distributions were collected from approximately 1,200,000 meteors observed at these two sources. The CMOR helion source is defined in the heliocentric ecliptic longitude $\lambda - \lambda_\odot$ and latitude $\beta$ as: $(\lambda - \lambda_\odot)_\mathrm{HE} = 337^\circ \pm 12^\circ$, $\beta_\mathrm{AH} = 2.4^\circ \pm 10^\circ$, whereas the anti-helion source has coordinates: $(\lambda - \lambda_\odot)_\mathrm{AH} = 202^\circ \pm 15^\circ$, $\beta_\mathrm{AH} = 2.4^\circ \pm 11^\circ$, where the apex of Earth's motion is at ($\lambda - \lambda_\odot = 270^\circ$).

The sensitivity of meteor radars is complex in nature \citep[e.g.][]{Janches_etal_2015} and involves multi-facility observations, extensive analysis, and numerous assumptions \citep{Weryk_Brown_2013}. For the purpose of this paper, we adopt the ionization limit used in \citet{CampbellBrown_2008} as
\begin{equation}
    I^\star = \frac{m_\mathrm{met}}{10^{-7}} \left( 
    \frac{
    \sqrt{v_\infty^2 + v_\mathrm{esc}^2}
    }
    {30 000} \right)^{3.5} = \frac{m_\mathrm{met}}{10^{-7}} \left( 
    \frac{
   v_{GF}
    }
    {30 000} \right)^{3.5},
    \label{EQ:Ionization}
\end{equation}
where the mass of the meteoroid $m_\mathrm{met}$, the relative hyperbolic excess impact velocity $v_\infty$, the escape velocity $v_\mathrm{esc}$, and the impact (gravitationally focused) velocity $v_{GF}$ are in SI (MKS) units, i.e, a meteoroid with $m_\mathrm{met} = 10^{-7}$ kg and $v_{GF} = 30,000$ m s$^{-1}$ has $I^\star = 1$. \citet{CampbellBrown_2008} estimated that $I^\star = 1.0$ is the detection limit for CMOR, which is the value we use in this paper. Note that \citet{CampbellBrown_2008} derived their expression using an extensive review done in \citet{Bronshten_1983} and our Eq.~\ref{EQ:Ionization} provides a simplified description of meteor ionization. The process depends on meteor velocity, composition, or atmospheric conditions \citep{Jones_1997} and is still an open topic of research \citep[e.g.,][]{Brown_Weryk_2020}.

In our model, we first calculate the orbital elements of each meteoroid $a,e,I$ using the heliocentric position and velocity vectors from the numerical simulation rather than the binned data stored in the 3D histogram. Then we calculate the collisional probability $\mathcal{C}_p$ of each meteoroid record with an idealized Earth ($a=1.0$ au, $e= 0.0$, $I=0.0^\circ$) using the \citet{Kessler_1981} method. 
\added{This method uses the orbital elements of the projectile and the target to analytically determine their collisional probability per unit time. For our idealized Earth this method is identical to that of \citep{Opik_1951}.} 
Multiplying $\mathcal{C}_p$ by the \replaced{number of meteoroid records in the simulation $N$}{size dependent model normalization parameter (Sec. \ref{SEC:SFD}} yields the number of impactors at Earth. Using the \replaced{model SFD}{particle size, SFD index, $v_{GF}$ and Eq. \ref{EQ:Ionization}}, we determine what fraction of meteoroids in the size bin impacting Earth is detectable by the radar; i.e., $I^\star \ge 1$. After investigating all meteoroid records in the numerical simulation, this procedure results in four histograms of $a$, $e$, $I$, and $v_\infty$ for meteoroids that would be detectable by CMOR, which we collectively summarize as our model measurement $\mathcal{M}_{O}$. The observed distribution of orbital elements from CMOR radar is then $\mathcal{O}_{O}$.

\subsubsection{The total cross-section of the Zodiacal Cloud}
\label{SEC:Constraint_4}
\label{SEC:Constriants_Cross}
\citet{Gaidos_1999} derived an effective emitting area for the Zodiacal Cloud of $\Sigma_\mathrm{ZC} = 1.12 \times 10^{17}$ m$^{2}$ by assuming that the dust particles emit blackbody radiation at 260 K \citep{Reach_etal_1996} and have the bolometric luminosity is $8 \times 10^{-8} L_\odot = 3.12 \times 10^{19}$ W \citep{Good_etal_1986}.
\citet{Nesvorny_etal_2010} used a dynamical model constrained by Infrared Astronomical Satellite (IRAS) observations of thermal emission to estimate $\Sigma_\mathrm{ZC} = 2.0 \pm 0.5 \times 10^{17}$ m$^{2}$.
\citet{Nesvorny_etal_2011JFC} estimated the total cross-section of their modeled Zodiacal Cloud to be $1.7 \times 10^{17} < \Sigma_\mathrm{ZC} < 3.4 \times 10^{17}$ m$^{2}$, in agreement with their previous estimate, but approximately twice the \citeauthor{Gaidos_1999} estimate.
%\citet{Nesvorny_etal_2010} discusses out that \citet{Gaidos_1999} underestimated this value by a factor of 2 due to , which pushes the estimate within both latter works uncertainty ranges. 
%Furthermore, \citet{Nesvorny_etal_2011JFC} showed that JFCs are the dominant population with respect to the Zodiacal Cloud cross-section, contributing more than 90\% of the total area, whereas the remaining meteoroid populations provide the broader and more diffused component of the ZC.
Unable to decide between these two estimates, we decided that for the purpose of this paper we will conservatively assume two scenarios: (A) $\Sigma_\mathrm{ZC} = 2.0 \times 10^{17}$ m$^{2}$, based on dynamical modeling works, and (B) $\Sigma_\mathrm{ZC} = 1.12 \times 10^{17}$ m$^{2}$ using \citet{Gaidos_1999}. In our model, we calculate the total cross-section of the Zodiacal Cloud by summing the cross-sections of all meteoroids in the model assuming they are perfect spheres; i.e, $ \mathcal{M}_\Sigma = \sum_i 0.25 \pi D^2_i$.
% \begin{equation}
%     \mathcal{M}_\Sigma = \sum_i 0.25 \pi D^2_i.
% \end{equation}

\subsubsection{Size-frequency distribution at Earth}
\label{SEC:SFD}
The interplanetary dust particles and sporadic meteoroids considered in our model range in size from $D=1~\mu$m up to $D=10,000~\mu$m. We exclude particles smaller than $1~\mu$m as a significant fraction of sub-micron particles that intercept the Earth are interstellar \citep{Strub_etal_2019}.  We place our upper limit at $10,000~\mu$m, or 1~cm, well below the lower size limit for asteroids (${\sim}1$~m) and approximately the size above which meteoroids are too scarce to be detected in large numbers by ground-based networks. Even within these limits, no single instrument can characterize the entire SFD of interplanetary dust particles and meteoroids.
%Meteoroids and dust arrive at Earth with a certain size-frequency distribution. However, no single instrument can characterize the entire distribution of interplanetary dust particles and meteoroids ranging in diameter from $D=1~\mu$m up to $D=10,000~\mu$m. Particles smaller than $D=1~\mu$m impacting Earth are significantly represented by interstellar grains \citep{Strub_etal_2019}, while meteoroids larger than $D=10,000~\mu$m are too scarce to be detected with good statistics by localized facilities around Earth, and too small to be detected by global networks \citep{Brown_etal_2002}. 
%Furthermore, since more than one abundant source of meteoroids feeds the meteoroid complex at Earth, we need the global size-frequency distribution to be dominated by JFC influx while keeping the remaining populations insignificant. Additionally, there are numerous meteor showers observed annually on both hemispheres \citep{Jenniskens_etal_2020,Brown_etal_2010,Pokorny_etal_2017_SAAMER}. 

Fortunately, at least for some range of meteoroid diameters the data is abundant enough to constrain the SFD at Earth.
%and confirm that their influx is dominated by JFCs. 
\citet{Pokorny_Brown_2016} determined the SFD of radar and optical meteors using more than 5,000,000 meteor echoes observed by CMOR and 3,106 optical meteors observed by CAMO (the Canadian Automated Meteor Observatory). Both samples were scrubbed of meteor showers to avoid contamination from meteoroid populations originating in singular sources \citep[for daily fluctuations of the SFD see Fig.~7 in][]{Pokorny_Brown_2016}. From these two independent observations, the authors determined that the SFD for meteors with masses $10^{-8}\mathrm{~kg} < m_\mathrm{met} < 10^{-4}\mathrm{~kg}$ has a differential mass index of $s = -2.1 \pm 0.08$. These values can be converted to a differential size-frequency index $\mathcal{S}=-4.3 \pm 0.24$ for the range of diameters $212.2\mathrm{~}\mu\mathrm{m} < D <  4571\mathrm{~}\mu\mathrm{m}$ assuming spherical particles and a bulk density of $\rho=$~2,000~kg m$^{-3}$. \citet{Galligan_Baggaley_2004}, using the AMOR radar, showed that $\mathcal{S}$ is shallower for smaller meteors and reported $\mathcal{S}=-4.09 \pm 0.03$ for meteors as small as $m_\mathrm{met}=10^{-10}$ kg (or $D=45.7~\mu$m). 

In our model, we fit the size-frequency distribution  \added{at the Earth} with a broken power law defined by two differential size indexes $a_1$ and $a_2$, the break point $\mu$, and the offset $b$ as:
%\begin{equation}
%    D>\mu, N(D) = 10^b D^{a_1}; D\le \mu, 10^b D^{a_2} \mu^{a_1-a_2}.
%    \label{EQ:Broken_Power_Law}
%\end{equation}
\begin{align}
    N(D) &= 10^b \times \begin{cases}
        D^{a_2 + 1} \mu^{a_1 - a_2} & \textrm{ for } D \le \mu \\
        D^{a_1 + 1} & \textrm{ for } D > \mu
    \end{cases}
    \label{EQ:Broken_Power_Law}
\end{align}
where $10^b$ is a normalization constant. The index $a_1$ defines the slope of the broken power law for particles with $D$ larger than the break point $\mu$, whereas $a_2$ represents the slope for the smaller particles in the SFD.
%For break points $\mu<212.2~\mu$m 
As break points $\mu>212.2~\mu$m have not been reported, we will constrain the value of $\mu$ to be less than $212.2~\mu$m. The value of $a_1$ is the observed differential SFD index for large particles -- $\mathcal{M}_\alpha = a_1$ -- which we will compare to the observed value $\mathcal{O}_\alpha = -4.3 \pm -0.24$ from \citet{Pokorny_Brown_2016}.

We are not aware of any recent observations that constrain the value of $a_2$. \citet{Fixsen_Dwek_2002} derived a value of $a_2 = -2.3\pm0.1$ and a SFD break point $\mu=30~\mu$m from DIRBE and FIRAS instruments on the COBE satellite that observed the spectrum of the Zodiacal Cloud. This value is not directly comparable with the SFD of meteoroids impacting Earth due to the different effect of gravitational focusing on smaller and larger particles. Therefore, we will only constrain the value of $a_1$ and discuss the SFD of particles with $D<\mu$ in our discussion section.

\subsubsection{Goodness-of-fit function for model constraints}
To quantify which of the model's free parameters provide the best representation of observed reality, we construct a goodness-of-fit function $\chi^2$. The combination of free parameters that minimizes $\chi^2$ will be considered our best fit. We define the total model $\chi^2_\mathrm{tot}$ as follows
\begin{equation}
    \chi^2_\mathrm{tot} = \chi^2_R + \chi^2_{\dot{M}} + \chi^2_O + \chi^2_\alpha,
    \label{EQ:Total_Likelihood}
\end{equation}
where $\chi^2_R$, $\chi^2_{\dot{M}}$, $\chi^2_O$, and $\chi^2_\alpha$ are the goodness-of-fit of the heliocentric distance profile, mass accretion rate at Earth, orbital distributions of meteors at Earth, and size-frequency distribution index. These goodness-of-fit functions are defined as follows:
\begin{equation}
    % \chi^2_R = \frac{(\mathcal{M}_R - \mathcal{O}_R)^2}{\sigma_R^2},
    % \chi^2_M = \frac{(\mathcal{M}_M - \mathcal{O}_M)^2}{\sigma_M^2},
    % \chi^2_\Sigma = \frac{(\mathcal{M}_\Sigma - \mathcal{O}_\Sigma)^2}{\sigma_\Sigma^2},
    % \chi^2_\alpha = \frac{(\mathcal{M}_\alpha - \mathcal{O}_\alpha)^2}{\sigma_\alpha^2},
    \chi^2_i = \frac{(\mathcal{M}_i - \mathcal{O}_i)^2}{\sigma_i^2},
\end{equation}
where $\mathcal{M}_i$ are the model values, $\mathcal{O}_i$ are the observed values, and $\sigma_i$ are the uncertainties associated with the observed values for $i \in \{R, \, \dot{M}, \, \Sigma, \, \alpha \}$. Unfortunately, the orbital element distributions from \citet{CampbellBrown_2008} do not come with uncertainty estimates. To calculate $\chi^2_O$, we adopt from \citet{CampbellBrown_2008} that from the total number of observed meteors $N_\mathrm{CMOR} = 2.35 \times 10^6$ approximately $P_\mathrm{HE/AH} = 51\%$ are in the helion/anti-helion source. \citet{CampbellBrown_2008} published distributions of four meteor orbital parameters: semimajor axis $a$, eccentricity $e$, inclination $I$, and the hyperbolic excess impact velocity $v_\infty$, each distributed into 100 uniformly spaced bins. All distributions in \citet{CampbellBrown_2008} are normalized, so we estimate the average variance in each bin as 
\begin{equation}
    \sigma^2_O = \left[\frac{N_\mathrm{CMOR} P_\mathrm{HE/AH}}{100}\right]^{-1/2} = 9.13\times 10^{-3}.
\end{equation}
Then we can estimate $\chi^2_O$ using following expression
\begin{equation}
    \chi^2_O = \sum_{i=1}^{400}\frac{(\mathcal{M}_{O_i} - \mathcal{O}_{O_i})^2}{400\sigma^2_O },
\end{equation}
where the factor 400 comes from the summation over 400 histogram bins. %The value of $\chi^2_O$ would be orders of magnitude higher than values of other goodness-of-fit functions in case of missing factor of 400 and $\chi^2_O$ would solely dominate the total goodness-of-fit $\chi^2_\mathrm{tot}$.
Note that we average over the 400 orbital element bins to assess the average chi-square for the distribution. Otherwise, $\chi^2_O$ would overwhelmingly dominate $\chi^2_\mathrm{tot}$ and the constraints offered by other observations would essentially be ignored.
There are other sources of uncertainties connected with meteor radar measurements such as the ionization efficiency variability, uncertainties in meteor velocity, and direction determination \citep[e.g.,][]{Mazur_etal_2020} that are currently beyond the scope of the paper. We ran simple tests by increasing and decreasing our $\sigma^2_O$ by a factor of 2 and found no significant changes in the results.

\subsubsection{Model parameters and minimization process}
Loading the JFC meteoroid model into our collisional grooming code distributes 371,247,921 unique records into 26 logarithmically scaled size bins. We perform a search for the minimum value of $\chi^2_\mathrm{tot}$ (defined in Eq.~\ref{EQ:Total_Likelihood}) over a four-parameter phase space $(\mathcal{P}_\alpha, \mathcal{P}_\gamma, \mathcal{P}_{A_s}, \mathcal{P}_{B_s})$  composed of: (1) the differential power law index for the SFD of the source population of JFCs meteoroids, $\mathcal{P}_\alpha$;~(2)~the JFC initial perihelion weighting parameter $\mathcal{P}_\gamma$, where each meteoroid stream carries weight $W={q^\star}^{\mathcal{P}_\gamma}$; (3) a parameter that determines the material property constant $A_s$, $\mathcal{P}_{A_s} = 100 A_s$ J kg$^{-1}$ \citep[where $A_s$ is the parameter in][after converting cgs to MKS; see Section \ref{SEC:Collision_Conditions}]{Krivov_etal_2005}; and (4) the material property constant $B_s = \mathcal{P}_{B_s}$ (see Eq.~\ref{EQ:Binding_Energy}). Parameters $\mathcal{P}_\alpha$ and $\mathcal{P}_\gamma$ control the initial size-frequency and heliocentric distance distribution in the particle cloud, whereas $\mathcal{P}_{A_s}$ and $\mathcal{P}_{B_s}$ control the collisional grooming process. We use a factor of 100 in $\mathcal{P}_{A_s} = 100 A_s$ to directly see the multiplication factor against values of $A_s$ used in \citet{Krivov_etal_2005} and \citet{Kuchner_Stark_2010}. For additional discussion of the influence of $\mathcal{P}_\gamma$ see Section 2.1 in \citet{Nesvorny_etal_2011JFC}.

The absolute scaling of the number of dust particles in the Zodiacal Cloud is determined by the initial value of the total Zodiacal Cloud cross-section $\Sigma_\textrm{ZC, init}$
for the collision-less cloud. In Scenario A, we select $\Sigma_\textrm{ZC, init} = 2.0 \times 10^{17}$~m$^{2}$, while for Scenario B we set $\Sigma_\textrm{ZC, init} = 1.12 \times 10^{17}$~m$^{2}$. As soon as the collisions are iteratively taken into account $\Sigma_\mathrm{ZC}$ decreases. Once a steady state is reached, the model has a total Zodiacal Cloud cross-section of  $\Sigma_\textrm{ZC, steady}$. The entire model is then re-scaled by a factor of $\Sigma_\textrm{ZC, init}/\Sigma_\textrm{ZC, steady} > 1.0$, and the model is collisionally groomed again until a second, augmented steady state is achieved. If the second augmented steady state is within 10\% of $\Sigma_\textrm{ZC, init}$ -- i.e., $| \ln (\Sigma_\textrm{ZC, init}/\Sigma_\textrm{ZC, steady}) | < 0.0953 $ -- then the collisional grooming is considered complete. If this condition is not satisfied, further rescalings are performed. A single rescaling is usually sufficient due to the tenuous nature of the current Zodiacal Cloud.

\subsection{Additional constraints and observations}
To keep this work concise, we include observational constrains that we are the most familiar with. Additional constraints for the shape of the Zodiacal Cloud and its Jupiter-family comet component are for example: the shape and brightness of the inner Zodiacal Cloud \citep{Hahn_etal_2002}; the Zodiacal Cloud emission spectrum and the size-frequency distribution break point at 14 and 32 $\mu m$ \citep{Fixsen_Dwek_2002}; micro-cratering records on the Moon \citep{Hoerz_etal_1975}; the Parker Solar Probe dust count record from 0.05 au to 1 au \citep{Malaspina_etal_2020}; or the spacecraft impact crater records from Pegasus and LDEF \citep{Moorhead_etal_2020}. We \added{compare our model results with the Parker Solar Probe in Sec. \ref{SEC:PSP_and_massLoss} and }leave the inclusion of more observational constrains for future work, where we will include additional major dust and meteoroid populations.

\section{Results}
In this section, we insert the entire JFC dynamical model into the collisional grooming code and look for the model realization that best fits reality as determined by the constraints described in Sec.~\ref{SEC:Constriants}.

\subsection{Constraining the fitting parameter space - Scenario A}
\label{SEC:1D_Analysis_ScenarioA}
\begin{figure}
\epsscale{1.0}
\plotone{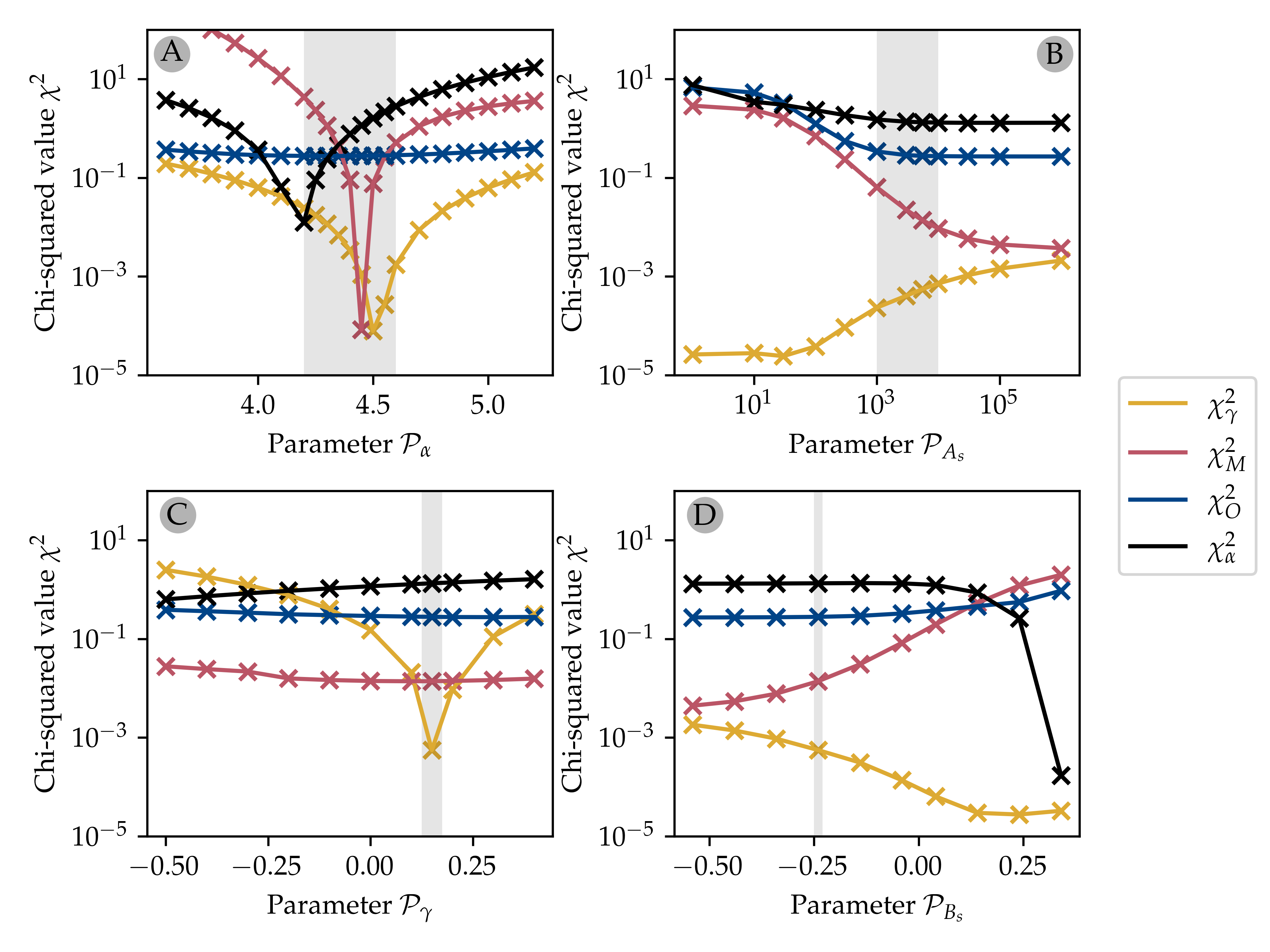}
\caption{\label{FIG:Parameter_Initial_Seach}
Goodness-of-fit function variations for different values of model parameters $\mathcal{P}$. Panel (A) shows the influence of $\mathcal{P}_\alpha$ on the goodness-of-fit functions for fixed $\mathcal{P}_\gamma = 0.15, \mathcal{P}_{A_s} = 5500, \mathcal{P}_{B_s} = -0.24$. Goodness-of-fit functions are color-coded: $\chi^2_R$ is shown in yellow, $\chi^2_{\dot{M}}$ in pink, $\chi^2_O$ in blue, and $\chi^2_\alpha$ in black. The gray region shows the parameter space we used for our focused fitting $\mathcal{P}_\alpha \in [4.2, 4.6]$. Panels (B-D) show the same as panel (A) but for parameters $\mathcal{P}_{A_s}, \mathcal{P}_{\gamma}, \mathcal{P}_{B_s}$, respectively. The focused fitting parameter space is (B) $\mathcal{P}_{A_s} \in [1000,8000]$, (C) $\mathcal{P}_\gamma \in [0.125, 0.175]$, and (D) $\mathcal{P}_{B_s} = -0.24$.
}
\end{figure}

Due to the computational demand of each model realization (6-8 hours of real time) and the broad parameter space, we first focused on determining the influence of each fitting parameter separately. We fix three of the four different fitting parameters $\mathcal{P}$ in our initial reconnaissance and let the fourth vary across a wide range of values. After experimenting with 256 different permutations of four free parameters we found a solution that provided a satisfactory total goodness-of-fit $\chi^2_\mathrm{tot}$ as well as partial goodness-of-fit values ($\chi^2_R$, $\chi^2_{\dot{M}}$, $\chi^2_O$, $\chi^2_\alpha$) for $\mathcal{P}_\alpha = 4.47$, $\mathcal{P}_\gamma = 0.15$, $\mathcal{P}_{A_s} = 5500$, and $\mathcal{P}_{B_s} = -0.24$. For the final round of our reconnaissance, we explore the effect of each individual parameter $\mathcal{P}$ while keeping the other parameters fixed at the previously mentioned values. Figure \ref{FIG:Parameter_Initial_Seach} shows variations in our four goodness-of-fit function values as a function of the four parameter values. Variations of $\chi^2$ values of individual parameters allow us to define a focused fitting parameter space (gray areas in Figure \ref{FIG:Parameter_Initial_Seach}). This parameter space constrains the ranges of all model parameters $\mathcal{P}$ and is used to fit all four model parameters at the same time.

%%%%
%%%

% filt_condition = ( (model_As == 5000) & (model_Bs == -0.24) & (model_gamma == 0.15))
% filt_condition = ( (model_alpha == 4.47) & (model_Bs == -0.24) & (model_As == 5500))
% filt_condition =( (model_alpha == 4.47) & (model_Bs == -0.24) & (model_gamma == 0.15))
% filt_condition = (  (model_alpha == 4.47) & (model_gamma == 0.15) & (model_As == 5500) )

\subsubsection{Influence of $\mathcal{P_{\alpha}}$}
First, we examine the influence of $\mathcal{P_{\alpha}}$, which represents the size-frequency distribution of particles at the source. Figure \ref{FIG:Parameter_Initial_Seach}A shows the most significant drop in $\chi^2_\mathrm{tot}$ for the differential size index at Earth ($\chi^2_\alpha$) and the JFC mass accreted at Earth ($\chi^2_{\dot{M}})$. These two functions go from indicating very poor fits ($\chi^2>10$) to indicating very good fits ($\chi^2<0.1$) inside the range we investigated ($\mathcal{P}_\alpha \in [3.6,5.3]$). The minimum values for these two goodness-of-fit functions are located between $\mathcal{P_{\alpha}} =4.20$ (for $\chi^2_\alpha$) and $\mathcal{P_{\alpha}} = 4.50$ (for $\chi^2_{\dot{M}}$). We also see a significant drop in $\chi^2_R$ but only in the log-scale since the value is very close to 0. The value itself does not go above $\chi^2_R > 0.2$ because we kept the value $\mathcal{P}_\gamma = 0.15$ which provides a good fit to the heliocentric distance distribution of JFC meteoroids in our model (Fig.~\ref{FIG:Parameter_Initial_Seach}C). The meteor orbit distribution is quite insensitive to $\mathcal{P}_\alpha$ as the radar detects meteors within a certain size range and thus, if the SFD is anchored at that size, changing the slope of the size-frequency distribution has a minimal effect on the distribution of detected meteors. Based on our initial analysis we select $\mathcal{P}_\alpha \in [4.2, 4.6]$ for our further analysis (gray region in Fig.~\ref{FIG:Parameter_Initial_Seach}A).

\subsubsection{Influence of $\mathcal{P_{A_\mathrm{s}}}$}
Parameter $\mathcal{P_{A_\mathrm{s}}}$ strongly influences the collisional lifetime of each particle by making particles more or less susceptible to collisions. Since one of our model constraints is the total cross-section of the currently observed Zodiacal Cloud (Section \ref{SEC:Constriants_Cross}), we can expect that $\mathcal{P_{A_\mathrm{s}}}$ mainly influences the total mass of particles impacting Earth, their orbital element distribution, and their size-frequency distribution. We do not expect this parameter to alter the heliocentric particle density profile significantly, because this profile is dominated by smaller particles that are evolving quicker than their larger counterparts. Figure \ref{FIG:Parameter_Initial_Seach}B shows our initial prediction, where we see that for $\mathcal{P_{A_\mathrm{s}}} < 10$ the goodness-of-fit functions are well-above $\chi^2>1$ but improve significantly as $\mathcal{P_{A_\mathrm{s}}}$ increases. Once $\mathcal{P_{A_\mathrm{s}}}$ reaches a value of 1000, the goodness-of-fit values start to plateau, showing that collisions in the particle cloud are less and less important above this threshold. Once reaching $\mathcal{P_{A_\mathrm{s}}} > 10,000$, the collisions in the cloud rarely serve to destroy particles, and our constraints cannot distinguish between different models. All models are equivalent to a collision-less model above this threshold. This means that for $\mathcal{P_{A_\mathrm{s}}} > 10,000$, particles in our model are not sensitive to collisions.  Based on our initial analysis we select $\mathcal{P}_{A_s} \in [1000, 8000]$ for our further analysis (gray region in \ref{FIG:Parameter_Initial_Seach}B). 

\subsubsection{Influence of $\mathcal{P_{\gamma}}$}
As suggested by its definition, the parameter $\mathcal{P_{\gamma}}$ most significantly influences the heliocentric distance profile of JFCs in the inner Zodiacal Cloud (Fig.~\ref{FIG:Parameter_Initial_Seach}C). Furthermore, this parameter changes the weighting of low-eccentricity and high-eccentricity orbits of meteoroids launched from JFCs because particles emitted near perihelion are ejected into more eccentric orbits. As a result, the eccentricity distribution of those meteoroids that impact Earth will vary with $\mathcal{P_{\gamma}}$. Despite this, Figure \ref{FIG:Parameter_Initial_Seach}C shows that $\mathcal{P_{\gamma}}$ has very little influence on $\chi^2_O$ and $\chi^2_{\dot{M}}$. This is due to the fact that $\chi^2_O$ and $\chi^2_M$ are quantities that are more sensitive to circularized meteoroids at 1 au and are accentuated by gravitational focusing at Earth. However, $\mathcal{P_{\gamma}}$ does, as expected, strongly influence the value of $\chi^2_R$, which has a minimum around $\mathcal{P_{\gamma}} = 0.15$. $\mathcal{P_{\gamma}}$ also influences the SFD of meteoroids impacting Earth and thus the value of $\chi^2_\alpha$. While the minimum of $\chi^2_\alpha$ lies around $\mathcal{P_{\gamma}} < -0.50$ for the default combination of parameters ($\mathcal{P}_\alpha = 4.47, \mathcal{P}_{A_s} = 5500, \mathcal{P}_{B_s} = -0.24$), we previously showed that $\chi^2_\alpha$ also strongly depends on $\mathcal{P_{\alpha}}$. We observe that increasing $\mathcal{P_{\gamma}}$ always steepens the SFD of the population of Earth's impactors. Since $\mathcal{P_{\gamma}}$ is the only parameter that strongly changes  $\chi^2_R$ we selected the interval $\mathcal{P_{\gamma}} \in [0.125,0.175]$ for our further investigation.

\subsubsection{Influence of $\mathcal{P_{B_\mathrm{s}}}$}
Parameter $\mathcal{P_{B_\mathrm{s}}}$ controls the scaling of the meteoroid binding energy  $E_\mathrm{bind}$ with its radius (Eq.~\ref{EQ:Binding_Energy}). $E_\mathrm{bind}$ is linearly proportional to $A_s$ and the meteoroid's mass, whereas $B_s$ determines the relationship between the binding energy and the radius of the target particle ($E_\mathrm{bind} \propto R_\mathrm{tar}^{B_s}$). For $B_s <0$ and meteoroid radii $<1$ meter (i.e., all meteoroids considered here), meteoroids are more difficult to break when compared to the scenario in which $B_s =0$, as their specific (per-unit-mass) binding energy increases. The fact that meteoroids are more resilient to fragmentation with decreasing size is something we would expect based on the latest in-situ observations of dust building blocks from \textit{ROSETTA} \citep{Guttler_etal_2019}. A suite of instruments on-board the \textit{ROSETTA} spacecraft observed dust particles and meteoroids of various sizes and morphologies and concluded that agglomerate dust particles are built-up from smaller sub-structures, with the smallest building block sizes commonly around $0.5-2.0~\mu$m. Apart from completely rock-solid dust grains, the aggregate dust particles tend to be more loosely packed together with increasing size and the smaller number of sub-structure connections decreases overall meteoroid strength.

Based on our initial 256 parameter permutation analysis (Sec. \ref{SEC:1D_Analysis_ScenarioA}) and the targeted follow-up analysis, we find that $\mathcal{P_{B_\mathrm{s}}}$ has little influence on the goodness-of-fit functions below $\mathcal{P_{B_\mathrm{s}}} < -0.10$ (see Fig.~\ref{FIG:Parameter_Initial_Seach}D). For $\mathcal{P_{B_\mathrm{s}}} > 0.0$, the size-frequency distribution started to show waviness, an oscillation of the SFD index between $D=10~\mu$m and  $D=1000~\mu$m. This results in a SFD that is impossible to fit with a broken power law  and invalid values of $\chi^2_\alpha$ for $\mathcal{P_{B_\mathrm{s}}} > 0.2$.
From this analysis, we conclude that $B_s$ does not have a strong effect on our goodness-of-fit as long as $\mathcal{P_{B_\mathrm{s}}} < -0.1$. Therefore we decided to use the original \citet{Krivov_etal_2006} value, $B_s = -0.24$, for the rest of the models in this article.

Concluding our one-dimensional parameter search, we constrain our parameter space for the best model fit search to the following phase space: $\mathcal{P}_{\alpha} \in [4.2,4.6]$; step: $0.05$, $\mathcal{P}_{A_s} \in [1000,8000]$; step: $1000$, $\mathcal{P}_\gamma \in [0.125, 0.175]$; step: $0.025$, and $\mathcal{P}_{B_s} = -0.24$, yielding 216 permutations in total.

\subsubsection{Finding the best fit}
\begin{figure}
\epsscale{1.0}
\plotone{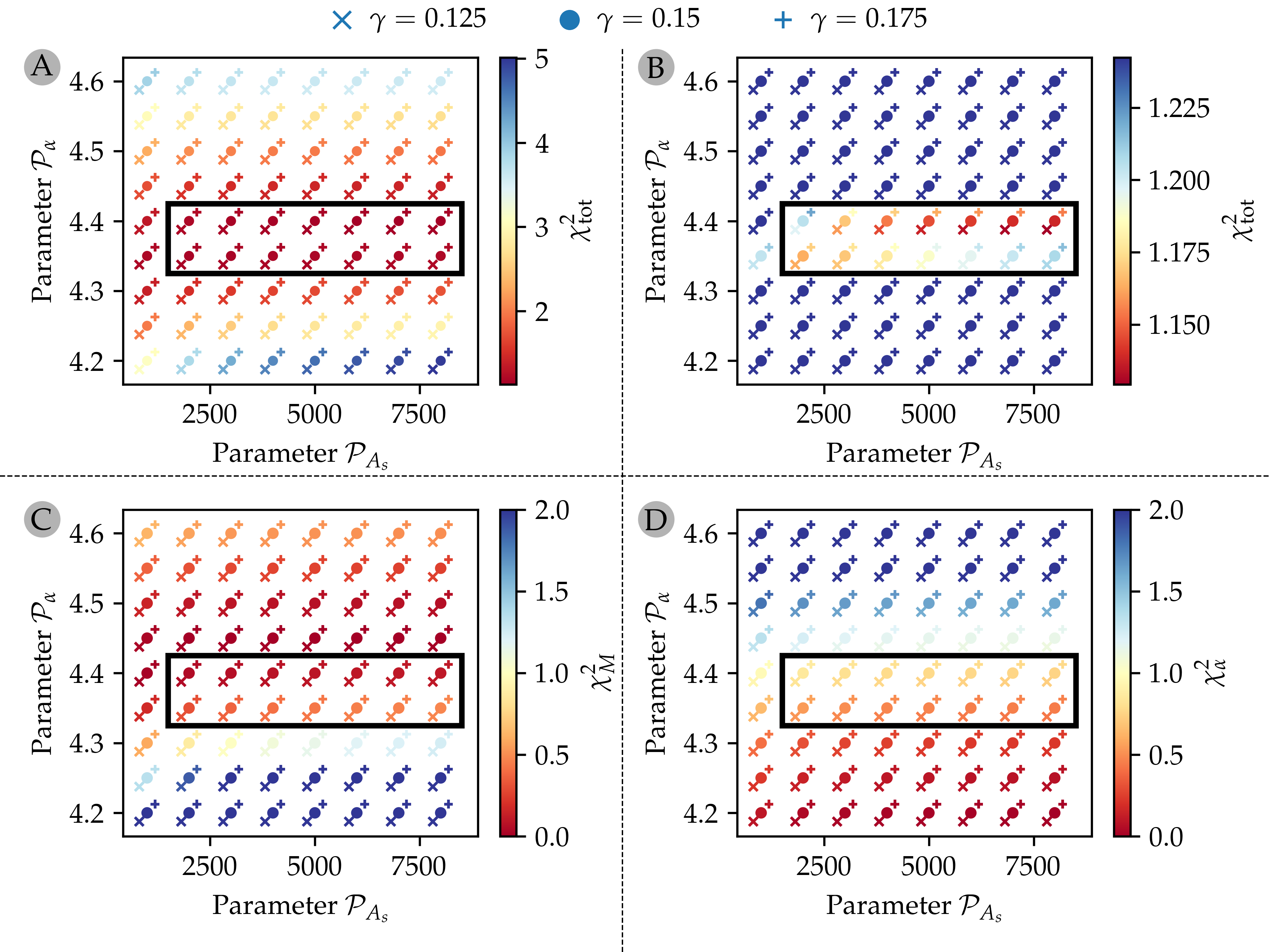}
\caption{\label{FIG:Parameter_Final_Search_Scenario_A}
Variations of the goodness-of-fit function $\chi^2_\mathrm{tot}$ of our Scenario A JFC model for 216 different parameters value combinations ($\mathcal{P_{\alpha}}$,$\mathcal{P_{A_\mathrm{s}}}$,$\mathcal{P_{\gamma}}$). The three symbols represent three different values of $\mathcal{P_{\gamma}}$: ($\times; 0.125$), ($\bullet; 0.15$), ($+; 0.175$). For clarity we offset $\times$ and $+$ symbols from $\bullet$ but all three symbols have the same values of ($\mathcal{P_{\alpha}}$,$\mathcal{P_{A_\mathrm{s}}}$). Each of the model realizations is color-coded by the value $\chi^2_\mathrm{tot}$ . The region of the best fitting parameters is denoted by the black rectangle:  $4.325 \le \mathcal{P_{\alpha}} \le 4.425$ and $1500 < \mathcal{P_{A_\mathrm{s}}} < 8500$. Panel (A) shows the full range of the total $\chi^2_\mathrm{tot}$ where we recorded the minimum value $\chi^2_\mathrm{tot} = 1.13$ for ($\mathcal{P_{\alpha}}$,$\mathcal{P_{A_\mathrm{s}}}$,$\mathcal{P_{B_\mathrm{s}}}$, $\mathcal{P_{\gamma}}$) = $(4.40,8000,-0.24, 0.125)$. Panel (B) contains the same information, but the color-scale now spans those values of $\chi^2_\mathrm{tot}$ within $10\%$ of the minimum value (1.13). Panel (C) shows the variation in the goodness-of-fit for mass accretion at Earth, $\chi^2_{\dot{M}}$, while panel (D) shows the variation in the goodness-of-fit for size-frequency distribution slope, $\chi^2_\alpha$.
}
\end{figure}

We ran 216 different model realizations within our fitting parameters constraints; the results are summarized in Figure \ref{FIG:Parameter_Final_Search_Scenario_A}A.  In order to display four-dimensional information in an image, we represent the parameter $\mathcal{P}_{\gamma}$ using three different symbols. Results of our focused $\chi^2_\mathrm{tot}$ minimization in Scenario A follow the results of our one-dimensional analysis described in Section \ref{SEC:1D_Analysis_ScenarioA}. Parameter $\mathcal{P}_{A_s}$, which scales the particle binding energy, produces small variations in $\chi^2_\mathrm{tot}$ for $\mathcal{P}_{A_s} > 2000$. The initial JFC differential size index $\mathcal{P}_{\alpha}$ minimizes $\chi^2_\mathrm{tot}$ around  $4.35 \le \mathcal{P_{\alpha}} \le 4.4$ (denoted by the black rectangle). The minimum value of $\chi^2_\mathrm{tot} = 1.13$ was recorded for ($\mathcal{P_{\alpha}}$,$\mathcal{P_{A_\mathrm{s}}}$,$\mathcal{P_{B_\mathrm{s}}}$, $\mathcal{P_{\gamma}}$) = $(4.40,8000,-0.24, 0.125)$, where all values in the black rectangle have  $\chi^2_\mathrm{tot} < 1.20$ and thus within $6\%$ of the minimum value. 

Figure \ref{FIG:Parameter_Final_Search_Scenario_A}B provides a closer look at the best fitting parameter combinations in our analysis. We constrained the color-scale to values with $\log_{10}\chi^2_\mathrm{tot} < 0.1$; i.e., $\chi^2_\mathrm{tot}<1.26$. We see that the best-fitting models have $\mathcal{P}_{A_s} > 6000$ and $\mathcal{P}_\alpha = 4.4$, where $\mathcal{P}_\gamma$ does not have a significant influence on the results. 

Our analysis shows that $\chi^2_\mathrm{tot} > 1$ for all parameters in our multi-dimensional parameter search, while in the one-dimensional analysis we were able to find parameter combinations for all four goodness-of-fit functions with values $\chi^2<0.1$, except for the radar orbit distribution $\chi^2_O$ that we show is only sensitive to $\mathcal{P}_{A_s}$. The reason for the low-quality fit is that the individual goodness-of-fit functions for mass accreted at Earth $\chi^2_{\dot{M}}$ and the SFD slope $\chi^2_\alpha$ do not show their minimum values for the same combinations of parameters (Fig.~\ref{FIG:Parameter_Final_Search_Scenario_A}C-D). 
%Such behavior shows a conflict between our constraints where any selection of parameters cannot satisfy all individual constraints. 
This suggests a potential conflict between our constraints that our JFC model might not be able to reproduce for any combination of model parameters.
For $\mathcal{P}_\alpha = 4.2$,  where we obtain the best fit for $\chi^2_\alpha$, the mass accreted at Earth is too high with $\dot{M}\approx 34,000$ kg day$^{-1}$, whereas for $\mathcal{P}_\alpha = 4.45$, where $\chi^2_{\dot{M}}$ is the lowest, the differential index $\alpha$ is too high with $a_1 \approx -4.56$.

\begin{figure}
\epsscale{1.2}
\plotone{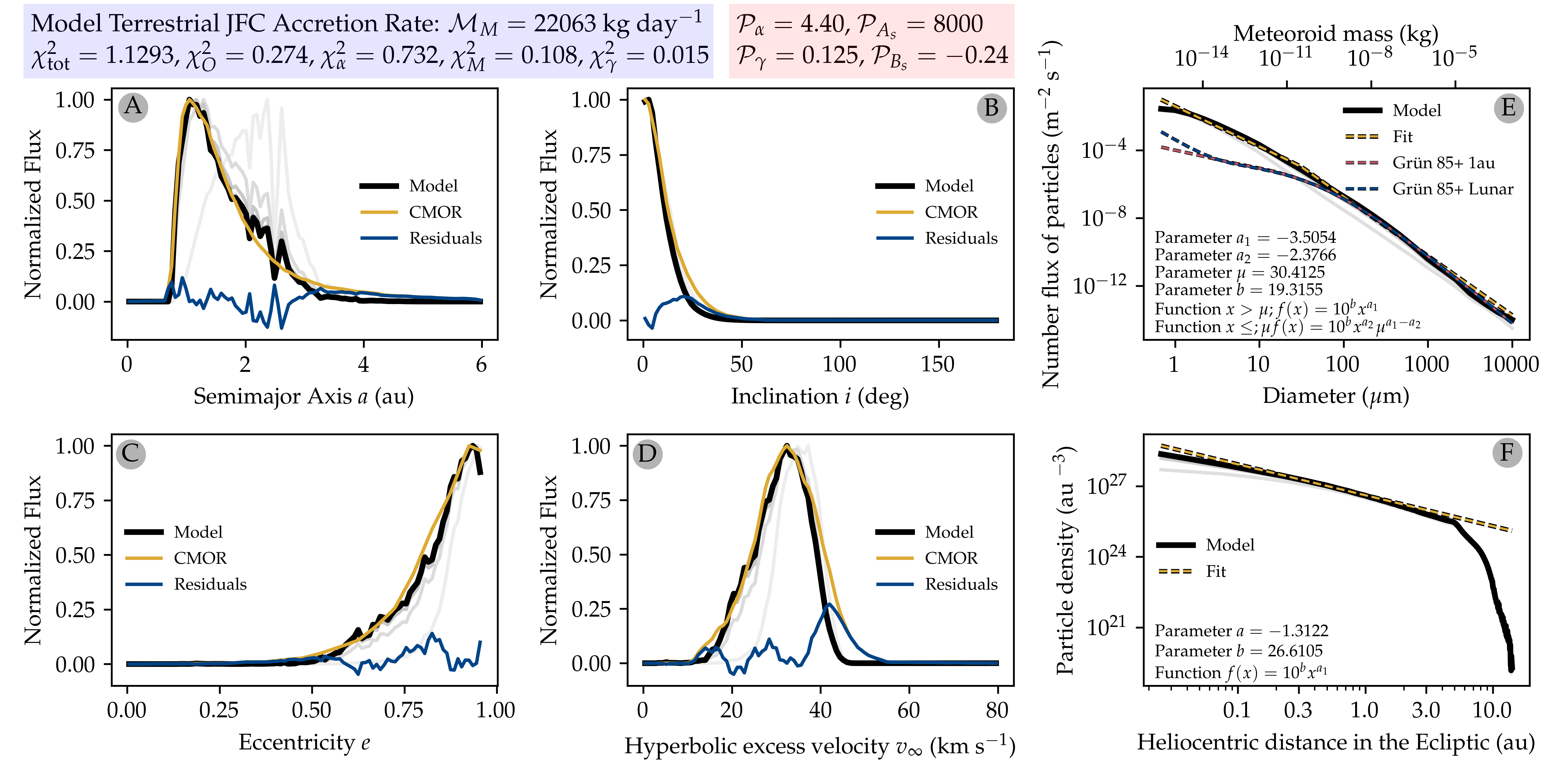}
\caption{\label{FIG:ScenarioA_Overview}
Overview of the best model for scenario A and its ability to fit our constraints. Panels (A-D) show the semimajor axis (A), inclination (B), eccentricity (C), and hyperbolic excess velocity (D) as observed by CMOR (orange solid line) and as reproduced in our model (black solid line). The gray solid lines show the collisional grooming iterations, where the lighter gray tones show the early iterations and the darker show more final ones. The blue line represents the model residuals. 
Panel (E) shows the size-frequency distribution of particles impacting Earth (black solid line), the broken-power law fit (orange dashed line) and \citet{Grun_etal_1985} interplanetary (pink dashed line) and lunar (blue dashed line) particle fluxes. The \citet{Grun_etal_1985} fluxes are multiplied by a factor of 3 to account for gravitational focusing effect that Earth experiences with respect to the Moon and interplanetary space \citep{Pokorny_etal_2018}. The legend shows broken power law fit parameters and the fitting function. All fluxes are cumulative.
Panel (F) shows the distribution of particle density with heliocentric distance measured in the ecliptic plane for our model (black solid line) and the fitted exponential (orange dashed line). The fit parameters and fitting function are shown in the legend.
Finally, the blue panel shows the value of $\mathcal{M}_{\dot{M}}$ and the goodness-of-fit function values. The red panel shows the model parameters $\mathcal{P}$.
}
\end{figure}

Now, we take a closer look at the ability of the best fit model to reproduce our constraints. Figure \ref{FIG:ScenarioA_Overview}A-D shows that our model fits the radar meteor orbit distribution from CMOR (orange solid lines) well with the exception of a lack of meteors faster than $v_\infty = 40$ km s$^{-1}$. The remaining orbital elements observed in helion and anti-helion sources are very close to model predictions (black solid lines). The observation-model residuals (blue solid line) show expected noise due to the radar orbit uncertainty that washes out the fine structures seen in our model. For example the radar data are not able to capture the effect of inner mean-motion resonances with Jupiter while the model shows a dip at 2.5~au corresponding to the 3:1 mean-motion resonance. The lack of faster meteors can be attributed to long-period comet populations such as the Halley-type comets not included in our model \citep{Pokorny_etal_2014} that are a part of the helion/anti-helion complex observed by CMOR.

The model cumulative size-frequency distribution shown in Fig.~\ref{FIG:ScenarioA_Overview}E (black solid line) spans over four orders of magnitude in particle diameter (or 12 orders of magnitude in particle mass). We fit the four-parameter broken power law (Eq.~\ref{EQ:Broken_Power_Law}) to this distribution (orange dashed line in Fig.~\ref{FIG:ScenarioA_Overview}E) where the fitting parameter $a_1$ represents the cumulative size index for the larger portion of particles. The observed differential size-frequency index $\mathcal{O}_\alpha=-4.3\pm 0.24$ is shallower than our best model predicts $\mathcal{M}_\alpha = a_1 = -4.505$, which is expected from our $\chi^2$ analysis shown in Fig.~\ref{FIG:Parameter_Final_Search_Scenario_A}. To match the observed values we would need the size-frequency distribution at source to be around $a_1 = -4.2$ (i.e. $\mathcal{P_\alpha} = 4.2)$. For comparison we also show the lunar (blue dashed line) and interplanetary (pink dashed line) cumulative particle flux as estimated in \citet{Grun_etal_1985}, where we multiplied the \citet{Grun_etal_1985} estimates by a factor of 3 to take into account the effect of gravitational focusing that JFC meteoroids experience at Earth \citep{Pokorny_etal_2018}. The gravitationally focused \citet{Grun_etal_1985} flux agrees well with our model between $80 <D < 10,000~\mu$m, where for smaller diameters our model predicts 1-2 orders of magnitude more particles than the \citet{Grun_etal_1985} model. This is due to our assumption of a single power law at the source, while there could be a break in the power law. We discuss the potential shortcomings of our single power law distribution in the Discussion section.

The variation in particle density with heliocentric distance close to the ecliptic (black solid line) shows a very good match to the observed power law scaling (orange dashed line; Fig.~\ref{FIG:ScenarioA_Overview}F). We fit the power law to our model values between 0.3 and 1.0 au. The model shows a decrease of the power law slope below $r_\mathrm{hel} = 0.1$ au due to the more frequent collisions in the innermost parts of the solar system. The rapid decrease in the particle density beyond 5.0 au is due to the initial distribution of JFCs. \added{The particle density beyond 5.0 au is not representative of the overall dust distribution due to contributions of outer solar dust populations originating from long-period comets and Edgeworth-Kuiper belt objects \citep{Poppe_2016}.} Note that the particle density in the innermost parts of the solar system ($r_\mathrm{hel}<0.1$ au) might not be correctly represented due to currently unknown effects on dust that are being reported by the Parker Solar Probe \citep{Stenborg_etal_2021} and that are not included in the model. 

In Scenario A, we use the normalization to the \citet{Nesvorny_etal_2010} and \citet{Nesvorny_etal_2011JFC} model values, $\Sigma_\mathrm{ZC} = 2.0 \times 10^{17}$ m$^{2}$ for the total cross-section of JFCs. Our results show that our model can either fit the observed mass accretion at Earth or the observed size-frequency distribution of meteoroids at Earth, but it fails to fit both constrains simultaneously. The radar meteor orbit distribution and the particle density variations with heliocentric distance are reproduced well for a variety of model parameters due to their insensitivity to parameter $\mathcal{P}_\alpha$ which determines the size-frequency distribution at the source. From our analysis, we see that in order to fit all constrains well we would need to decrease the mass accreted at Earth by a factor of 2. In Section \ref{SEC:Constriants_Cross} we consider two values for $\Sigma_\mathrm{ZC}$: Scenario A: $\Sigma_\mathrm{ZC} = 2.0\pm 0.5 \times 10^{17}$ m$^{2}$; Scenario B: $\Sigma_\mathrm{ZC} = 1.12 \times 10^{17}$ m$^{2}$, where their ratio is 1.79; a value very close to a factor of 2.
%which close to a factor by which \citet{Gaidos_1999}  value (Scenario B) differs from the value used in Scenario A model estimates. 
We could also challenge the \citet{CarrilloSanchez_etal_2020} JFC mass flux and use the value estimated in \citet{CarrilloSanchez_etal_2016} who estimated $\mathcal{O}_{\dot{M}} = 34 600 \pm 13800$ kg day$^{-1}$. Such a modified value allows for a very good fit at $\alpha = -4.2$ and thus reconciling the inability to fit both the size-frequency distribution and the mass accretion rate at Earth at the same time. Using $\mathcal{O}_{\dot{M}} = 34 600 \pm 13800$ kg day$^{-1}$ we obtain the best fit for:  $\mathcal{P}_\alpha = 4.20, \mathcal{P}_\gamma = 0.175, \mathcal{P}_{A_s} = 8000, \mathcal{P}_{B_s} = -0.24$, with the goodness-of-fit value $\chi^2_\mathrm{tot}=0.298$, indicating a very good fit to all constraints.
%%%%
%%%

%%%%%%%%%%%%%%
%%%%%%%%%%%%%%
%%%% ALTHEA'S STAMINA VANISHED APPROXIMATELY HERE 
%%%% TO DO: BREW SOME POTION TO RESTORE ALTHEA'S STAMINA
%%%%%%%%%%%%%%
%%%%%%%%%%%%%%

%%%%
%%%

\subsection{Results - exploring Scenario B} 
\label{SEC:Explore_SB}
Motivated by results from Scenario A, we now change the model total Zodiacal Cloud cross-section estimate to that of \citet{Gaidos_1999}, i.e. $\Sigma_\mathrm{ZC} = 1.12 \times 10^{17}$ m$^{2}$. This will effectively decrease the amount of particles in our model to $56\%$ of our first model in Scenario A and therefore decrease the collision rates between meteoroids by the same amount. This change should result in a decreased mass flux at Earth for the same model parameters and longer collisional lifetimes of all particles.

\subsubsection{Scenario B: Constraining the fitting parameter space}
\begin{figure}
\epsscale{1.0}
\plotone{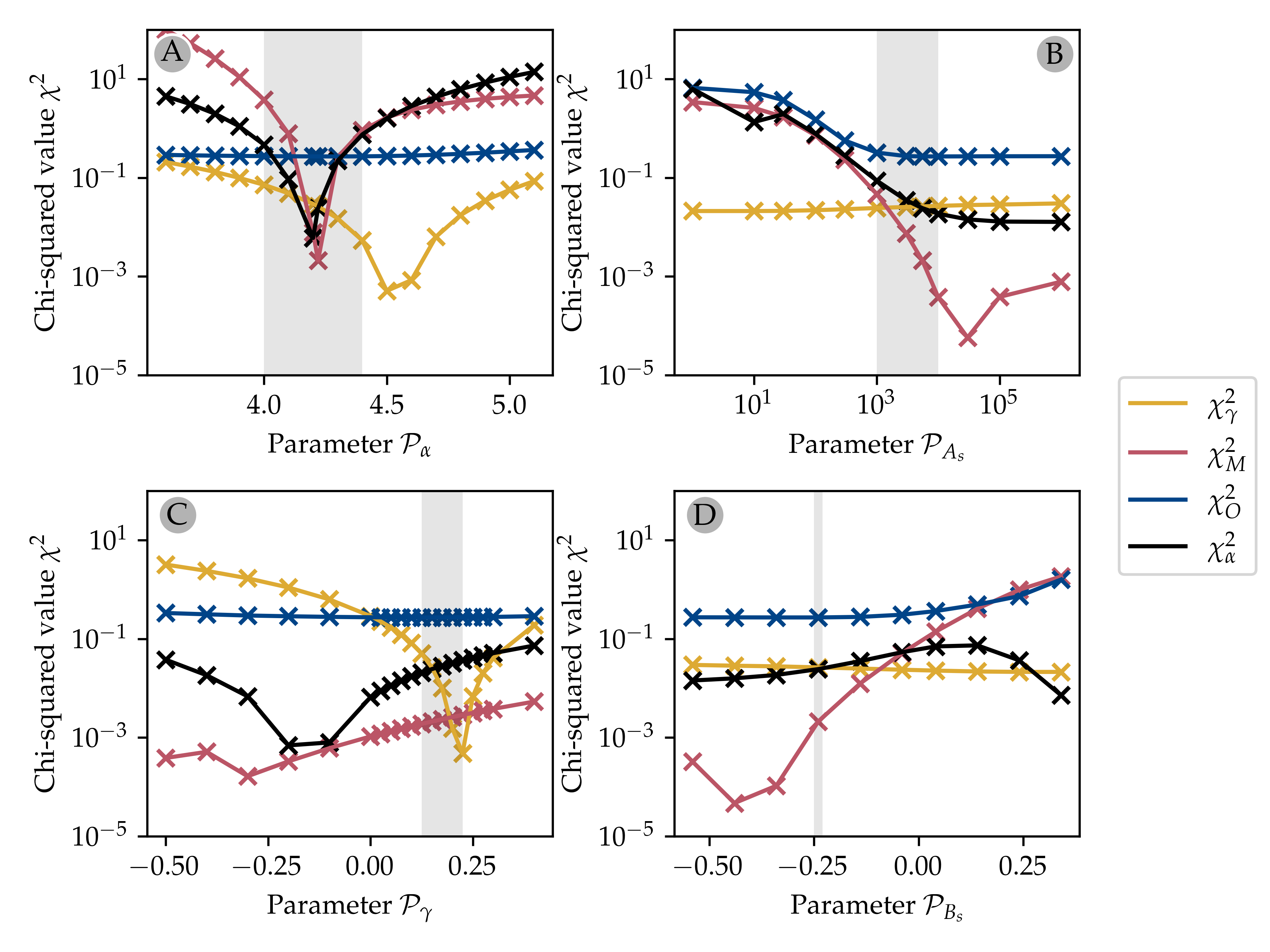}
\caption{\label{FIG:Parameter_Initial_Seach_SB}
The same as Fig. \ref{FIG:Parameter_Initial_Seach} but now using \citet{Gaidos_1999} total Zodiacal Cloud cross-section $\Sigma_\mathrm{ZC} = 1.12 \times 10^{17}$ m$^{2}$ for our JFC model. The focused fitting parameter space for Scenario B is: (A) $\mathcal{P}_{\alpha} \in [4.0,4.4]$ (B) $\mathcal{P}_{A_s} \in [1000,8000]$, (C) $\mathcal{P}_\gamma \in [0.125, 0.225]$, and (D) $\mathcal{P}_{B_s} = -0.24$.
}
\end{figure}

We perform the same initial parameter space analysis as in Scenario A by fixing three of our fitting parameters and exploring the remaining fitting parameter for a wide range of values (Fig. \ref{FIG:Parameter_Initial_Seach_SB}). In order to reflect lessons learned in Scenario A, the fixed parameter values are the following:  $\mathcal{P}_\alpha = 4.22, \mathcal{P}_\gamma = 0.15, \mathcal{P}_{A_s} = 5500, \mathcal{P}_{B_s} = -0.24$.

A significant difference compared to Scenario A is a much better agreement of minimum values of $\chi^2_{\dot{M}}$ and $\chi^2_\alpha$ for similar values of parameter $\mathcal{P}_\alpha$ in Fig. \ref{FIG:Parameter_Initial_Seach_SB}A. Both of these goodness-of-fit functions show a global minimum around $\mathcal{P}_\alpha = 4.2$, showing a good sign that the reduced amount of particles in the cloud can provide a much better fit to observed constraints. As expected, the good fits are found for smaller values of parameter $\mathcal{P}_{A_s}$ which does not show significant improvement beyond $\mathcal{P}_{A_s} > 3000$ (Fig. \ref{FIG:Parameter_Initial_Seach_SB}B). Interestingly, $\chi^2_R$ has a minimum for slightly larger value of $\mathcal{P}_\gamma$ than in Scenario A, which is correlated with the shift of the fixed parameter to $\mathcal{P}_\alpha = 4.22$ from $\mathcal{P}_\alpha = 4.47$ in Scenario A (Fig. \ref{FIG:Parameter_Initial_Seach_SB}C). Parameter $\mathcal{P}_{B_s}$ shows a worsening for all $\chi^2$ functions for $\mathcal{P}_{B_s}>-0.10$, thus we keep it at $\mathcal{P}_{B_s} = -0.24$ using the same value as in Scenario A.

Our parameter space for the best model fit search is thus following: $\mathcal{P}_{\alpha} \in [4.0,4.4]$; step: $0.05$, $\mathcal{P}_{A_s} \in [1000,8000]$; step: $1000$, $\mathcal{P}_\gamma \in [0.125, 0.225]$; step: $0.025$, and $\mathcal{P}_{B_s} = -0.24$, which yields 360 permutations in total.
%%%%
%%%

\subsubsection{Scenario B: The best model fit}
\begin{figure}
\epsscale{1.0}
\plotone{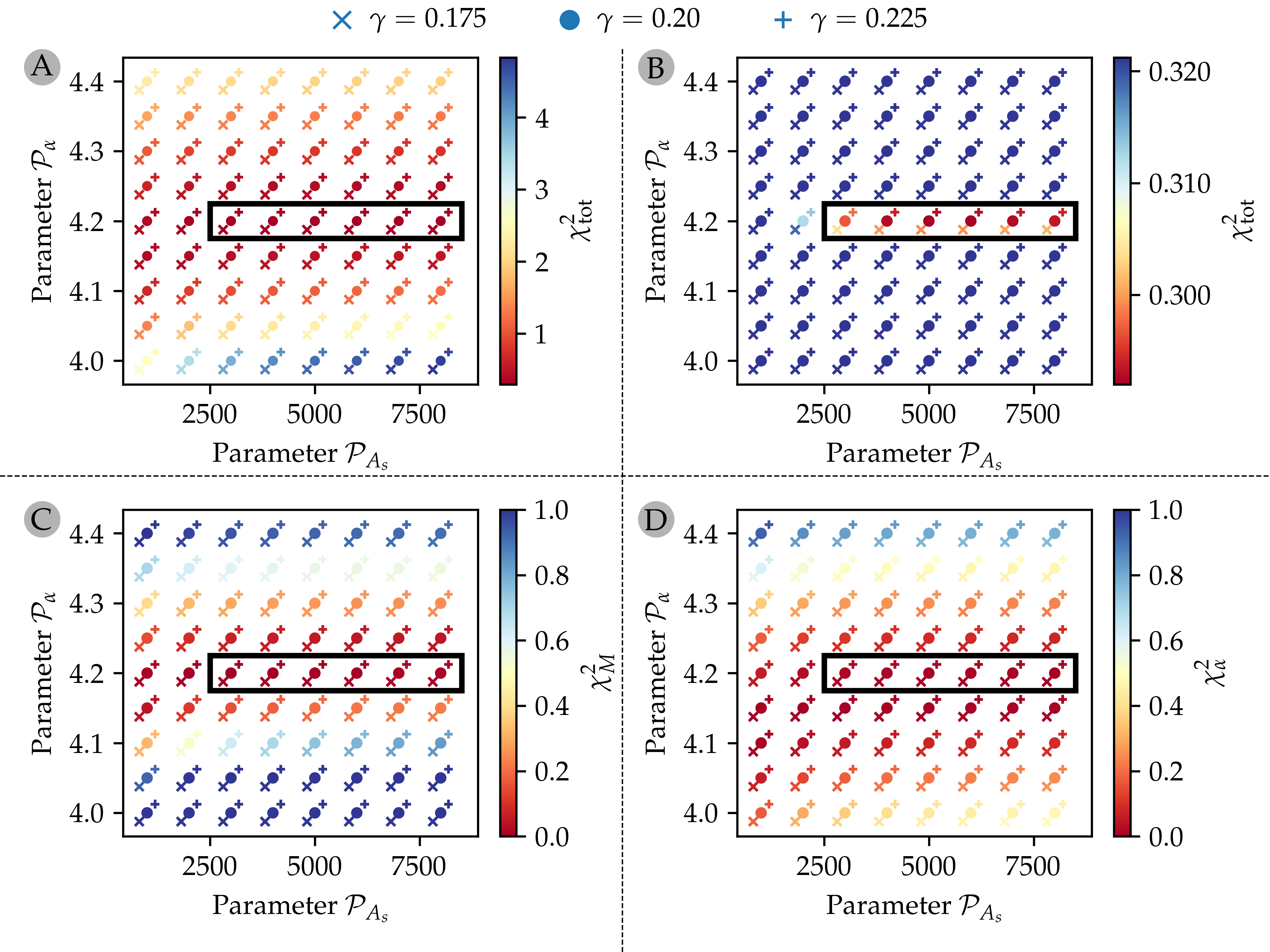}
\caption{\label{FIG:Parameter_Final_Search_Scenario_B}
{The same as Fig. \ref{FIG:Parameter_Final_Search_Scenario_A}} but now for Scenario B using the \citet{Gaidos_1999} total Zodiacal Cloud cross-section $\Sigma_\mathrm{ZC} = 1.12 \times 10^{17}$ m$^{2}$ for our JFC model. The three symbols represent three different values of $\mathcal{P_{\gamma}}$: ($\times; 0.175$), ($\bullet; 0.20$), ($+; 0.225$). For clarity, we offset $\times$ and $+$ symbols from $\bullet$ but all three symbols have the same values of ($\mathcal{P_{\alpha}}$,$\mathcal{P_{A_\mathrm{s}}}$). Each of the model realizations is color-coded by the value $\chi^2_\mathrm{tot}$. 
The region of the best fitting parameters is denoted by the black rectangle:  $4.1725 \le \mathcal{P_{\alpha}} \le 4.225$ and $1500 < \mathcal{P_{A_\mathrm{s}}} < 8500$. Panel (A) shows the full range of the total $\chi^2_\mathrm{tot}$ where we recorded the minimum value $\chi^2_\mathrm{tot} = 0.299$ for ($\mathcal{P_{\alpha}}$,$\mathcal{P_{A_\mathrm{s}}}$,$\mathcal{P_{B_\mathrm{s}}}$, $\mathcal{P_{\gamma}}$) = $(4.20,5000,-0.24, 0.20)$. Panel (B) shows the values within $10\%$ of the minimum value $\chi^2_\mathrm{tot}=0.292$. Panel (C) shows the variations of the goodness-of-fit for mass accretion at Earth $\chi^2_{\dot{M}}$, while panel (D) shows the same for the goodness-of-fit $\chi^2_\alpha$ for size-frequency distribution slope $\alpha$.
}
\end{figure}

For Scenario B, we ran 360 different model realizations using the parameter space defined in Sec. \ref{SEC:Explore_SB}. Figure \ref{FIG:Parameter_Final_Search_Scenario_B} shows the results of our search. Compared to Scenario A, the total goodness-of-fit is much better with the minimum $\chi^2_\mathrm{tot} = 0.292$ for ($\mathcal{P_{\alpha}}$,$\mathcal{P_{A_\mathrm{s}}}$,$\mathcal{P_{B_\mathrm{s}}}$, $\mathcal{P_{\gamma}}$) = $(4.20, 5000, -0.24, 0.20)$ than that in our Scenario A analysis ($\chi^2_\mathrm{tot} = 1.13$). This is due to the fact that both the size-frequency distribution fit $\chi^2_\alpha$ and the mass accreted at Earth fit $\chi^2_{\dot{M}}$ have minima for the same location in our four-dimensional parameter space (Fig \ref{FIG:Parameter_Final_Search_Scenario_B}C-D). For $\mathcal{P_{\alpha}} = 4.20$ and $\mathcal{P_{A_\mathrm{s}}}>3000$ both goodness-of-fit functions ($\chi^2_\alpha$,$\chi^2_{\dot{M}}$)  are $<0.015$, indicating a very good fit to the observed constraints. Since both $\chi^2_\alpha$ and $\chi^2_{\dot{M}}$ are very close to zero, the value of $\chi^2_\mathrm{tot}$ is dominated by the goodness-of-fit to radar meteor orbital distributions with $\chi^2_O > 0.27$ for all parameter permutations in our model. Note, that the quality of our best fit is similar to that found in Scenario A using the \citet{CarrilloSanchez_etal_2016} value for the JFC mass flux at Earth. Both of these best fits are achieved by using almost identical model parameters, where $\mathcal{P}_\alpha = 4.20$, $\mathcal{P}_{A_\mathrm{s}} > 3000$, $0.175<\mathcal{P}_{\gamma}<0.20$.

\begin{figure}
\epsscale{1.2}
\plotone{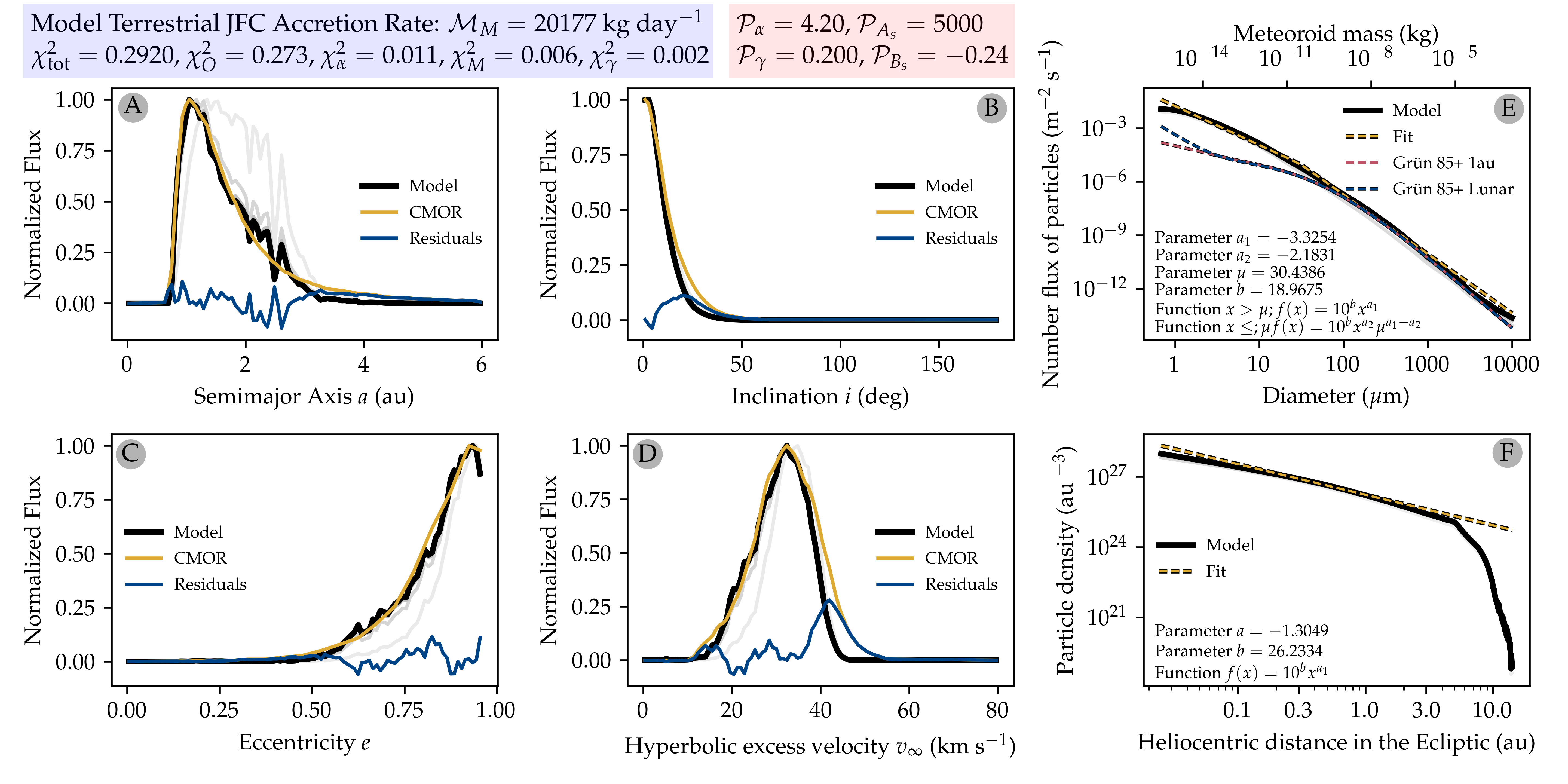}
\caption{\label{FIG:ScenarioB_Overview}
The same as Fig. \ref{FIG:ScenarioA_Overview} but now for the best model for Scenario B.}
\end{figure}

Figure \ref{FIG:ScenarioB_Overview} shows the ability of our best fitting model to reproduce our observational constraints. From individual $\chi^2$ values, we see that the Scenario B model provides a better fit in all analyzed aspects compared to the best fitting model in Scenario A. While the fit to meteor radar data and the particle density variations with the heliocentric distance is comparable to values recorded in Scenario A, the fits to the mass accreted at Earth and the size-frequency distribution index are reproduced very well with $\chi^2 \sim 0.01$.
The shallower size index also provides a better match to \citet{Grun_etal_1985} for micron-sized meteoroids (in Scenario A the number flux at Earth was 2-3 orders of magnitude higher in our model compared to that in \citet{Grun_etal_1985} for $D<5~\mu$m). 

%%%%
%%%

%%%%
%%%

\subsection{Summary of both fitting Scenarios}
To summarize our model minimization analysis, we find that Scenario A provides a low quality fit $\chi^2_\mathrm{tot} = 1.13$ assuming the \citet{CarrilloSanchez_etal_2020} JFC mass flux at Earth. The model is able to either reproduce the size-frequency distribution at Earth for $\mathcal{P}_\alpha = 4.2$ but provides ${\sim}2\times$ larger mass flux of JFC meteoroids at Earth, or reproduces well the mass flux for $\mathcal{P}_\alpha = 4.45$ but then provides a too steep meteoroid SFD at Earth. This discrepancy can be reconciled by using the \citet{CarrilloSanchez_etal_2016} JFC mass flux estimate that provides a much better fit to all constraints and results in the total goodness-of-fit value $\chi^2_\mathrm{tot} = 0.299$ for the parameter combination ($\mathcal{P_{\alpha}}$,$\mathcal{P_{A_\mathrm{s}}}$,$\mathcal{P_{B_\mathrm{s}}}$, $\mathcal{P_{\gamma}}$) = $(4.20,8000,-0.24, 0.175)$.

In Scenario B, we assume the total cross-section of meteoroids is 44\% smaller than in Scenario A following the \citet{Gaidos_1999} estimate. In this case, we find that our model can fit all constraints  presented here with $\chi^2_\mathrm{tot} = 0.292$ for the parameter combination ($\mathcal{P_{\alpha}}$,$\mathcal{P_{A_\mathrm{s}}}$,$\mathcal{P_{B_\mathrm{s}}}$, $\mathcal{P_{\gamma}}$) = $(4.20,5000,-0.24, 0.20)$. Should we assume that the \citet{CarrilloSanchez_etal_2016} JFC mass flux is the correct value, Scenario B suffers from similar problems we encountered in Scenario A. 

From our parameter search, we conclude that our best model fit is for the following parameters: $\mathcal{P_{\alpha}} = 4.20 \pm 0.10$, $\mathcal{P_{A_\mathrm{s}}}= 5000 \pm 4000$, $\mathcal{P_{B_\mathrm{s}}} = -0.24$, and $\mathcal{P_{\gamma}}= 0.20 \pm 0.025$. \added{We estimate the $1-\sigma$ interval for each parameter as the range where all individual $\chi^2$ values are below unity.} This means that Jupiter-family comet meteoroids are released from their source population with an initial differential size-frequency index $\alpha = -4.20 \pm 0.10$, their binding energy is $E_\mathrm{bind} = 5\times10^5 \pm 4\times10^5 R_\mathrm{met}^{-0.24} m_\mathrm{met}$, and their weighting follows the trend ${q^\star}^{0.20}$.

\section{Collisional lifetimes}
Having found the best fit to our constraints in Scenarios A and B, we can now calculate the collisional lifetimes for particles of various sizes and orbits in our model and compare them to other estimates used in recent Zodiacal Cloud models. In this Section, we compare the collisional lifetime $T_\mathrm{coll}$ (Eq.~\ref{EQ:Collisional_Lifetime}) obtained from our Zodiacal Cloud model to estimates from \citet{Steel_Elford_1986} and \citet{Grun_etal_1985}: $T_\mathrm{coll}\mathrm{(SE86)}$ and  $T_\mathrm{coll}\mathrm{(G85)}$.
Each of these estimates has been used in numerous dynamical modeling papers. $T_\mathrm{coll}\mathrm{(SE86)}$ was used for example in \citet{Wiegert_etal_2009}, \citet{Pokorny_etal_2014}, \citet{Pokorny_etal_2018}, \citet{Pokorny_etal_2019}, and \citet{Yang_Ishiguro_2018} while $T_\mathrm{coll}\mathrm{(G85)}$ was implemented in \citet{Nesvorny_etal_2010}, \citet{Nesvorny_etal_2011JFC}, \citet{Soja_etal_2019}, and \citet{Yang_Ishiguro_2018}. 

\begin{figure}
\epsscale{1.2}
\plotone{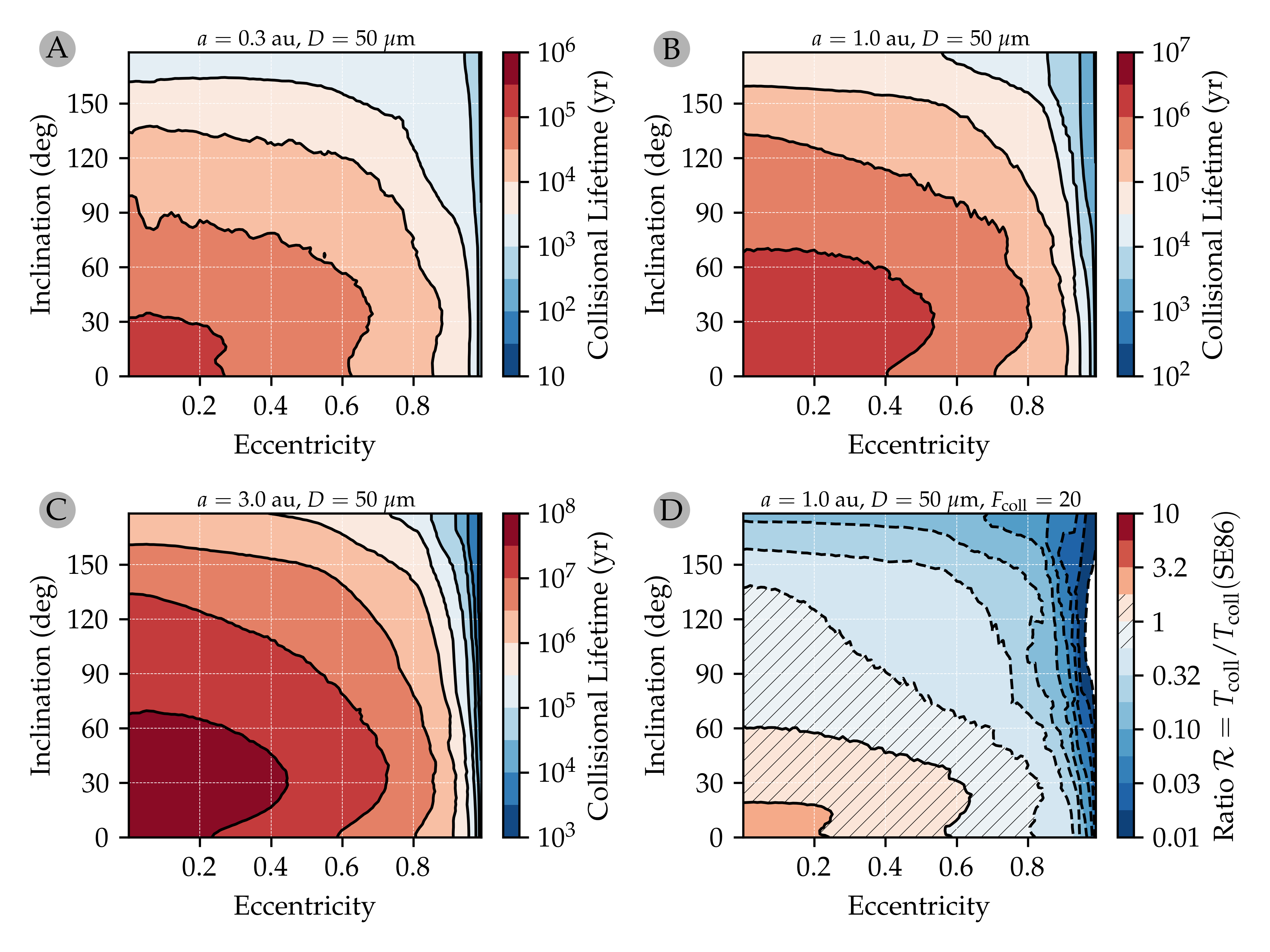}
\caption{\label{FIG:Collisions_Comparison}
Scenario B collisional lifetimes for $D=50~\mu$m meteoroid. Panels (A)-(C) show the collisional lifetime in years for the full range of bound orbits; eccentricity $0<e<1$, inclination $0<I<180^\circ$ where the color coding represents logarithmically spaced collisional lifetime. Panel (A) is for a particle with semimajor axis $a=0.3$ au, panel (B) for $a=1.0$ au, and panel (C) for $a=3.0$ au. Contours show boundaries between color-coded logarithmic steps. Panel (D) shows the ratio between collisional lifetimes in our model $T_\mathrm{coll}$ and $T_\mathrm{coll}\mathrm{(SE86)}$ with $\mathcal{F}_\mathrm{coll} = 20$ used in e.g., \citet{Pokorny_etal_2014, Pokorny_etal_2018}, i.e. the color-coding shows $\mathcal{R}=T_\mathrm{coll}/T_\mathrm{coll}\mathrm{(SE86)}$. The hatching shows areas within a factor of 1.78 of unity, i.e $\mathcal{R}=0.56-1.78$. Our model provides ${\sim}3\times$ longer collisional lifetimes for meteoroids on low-$e$ and low-$I$ orbits and 1-2 orders of magnitude shorter collisional lifetimes for high-$e$ and high-$I$ orbits (dark blue regions).
}
\end{figure}

Figure \ref{FIG:Collisions_Comparison} shows collisional lifetimes calculated in our best fit model in Scenario B using Eq.~\ref{EQ:Collisional_Lifetime} for a meteoroid with diameter $D=50~\mu$m assuming three different semimajor axes: $a=0.3$ au (Fig. \ref{FIG:Collisions_Comparison}A), $a=1.0$ au (Fig. \ref{FIG:Collisions_Comparison}B), and $a=3.0$ au (Fig. \ref{FIG:Collisions_Comparison}C). The collisional lifetime was calculated by averaging the collisional lifetimes over the full range of the meteoroid's argument of pericenter $\omega$, the longitude of the ascending node $\Omega$, and the mean anomaly $M$, where for each of the three variables we took 100 values in the range $(0,2\pi)$, i.e. using an average of $10^6$ different points in the phase-space $(\omega,\Omega,M)$. We calculated the collisional lifetime in our model for the full range of possible bound orbits, i.e. eccentricity $0<e<1$, and inclination $0<I<180^\circ$. 

%The collisional lifetimes follow expected results. Meteoroids on low-$e$ and low-$I$ orbits have the longest $T_\mathrm{coll}$ and with increasing $e$ and $I$ the collisional lifetime decreases. 
The collisional lifetimes in our model range over 5 orders of magnitude between the longest-lived particles and the shortest-lived ones, which are at retrograde and nearly parabolic orbits. Meteoroids on low-$e$ and low-$I$ orbits have the longest $T_\mathrm{coll}$ and the collisional lifetime decreases with increasing $e$ and $I$.
This behavior appears at all semimajor axes. 

Perhaps these variations in collisional lifetime are to be expected. The circular coplanar meteoroids orbit together with the bulk of the Zodiacal Cloud and therefore experience the smallest rates of destructive collisions due to very small relative impact velocities. On the other hand, retrograde meteoroids orbit against the flow and have much higher relative impact velocities. With increasing eccentricity, meteoroids penetrate into denser regions of the Zodiacal Cloud which naturally decreases their collisional lifetimes via increased rates of collisions. Moreover, their relative impact velocities are additionally increased due to higher orbital velocities closer to the Sun.

We calculated $T_\mathrm{coll}\mathrm{(SE86)}$ for the same meteoroids shown in Fig. \ref{FIG:Collisions_Comparison}B ($a=1.0$ au, $D=50~\mu$m) using the description given in \citet{Pokorny_etal_2018} and using the collisional lifetime multiplier $F_\mathrm{coll}=20$ that was determined in \citet{Pokorny_etal_2014} for HTC meteoroids and subsequently used in numerous works. In Figure \ref{FIG:Collisions_Comparison}D, we show the ratio of our model and $T_\mathrm{coll}\mathrm{(SE86)}$ collisional lifetimes $\mathcal{R} = T_\mathrm{coll}/ T_\mathrm{coll}\mathrm{(SE86)}$. While there is some common ground (hashed areas in Fig. \ref{FIG:Collisions_Comparison}D show $0.56<\mathcal{R}<1.78$), our model predicts ${\sim}10\times$ longer collisional lifetimes for low-$e$ and low-$I$ meteoroids, while for meteoroids on high-$e$ orbits our model predicts 1-2 orders of magnitude shorter collisional lifetimes than those estimated by $T_\mathrm{coll}\mathrm{(SE86)}$. These discrepancies are a natural consequence of assumptions made to calculate $T_\mathrm{coll}\mathrm{(SE86)}$.  \citet{Steel_Elford_1986} assume the relative impact velocity is constant, which leads to an underestimation of destructive collisions near the Sun while overestimating the destructive collision rates for meteoroids on similar orbits. 

\citet{Grun_etal_1985} does not take into account the particle inclination and makes various assumptions for the velocity and density scaling in the solar system. In Figure \ref{FIG:ScenarioB_Overview}E, we show that the \citet{Grun_etal_1985} and our model agree in predicted fluxes at 1 au for particles with $D>80~\mu$m but show 1-2 orders of magnitude difference for smaller grains. Additionally, in \citet{Grun_etal_1985} the relative impact velocity of dust particles in the Zodiacal Cloud scales only with the heliocentric distance and ignores particle eccentricity and inclinations. This disagreement in predicted flux at 1 au is likely caused by a more complex size-frequency distribution at the source than the one assumed here. If we introduced a broken power law with a break point below $D=30~\mu$m as suggested by \citet{Fixsen_Dwek_2002} and assumed a shallower power law index, this discrepancy would decrease.

Let us ignore for the moment the different flux and velocity distributions seen in our models and \citet{Grun_etal_1985}.  Eq.~\ref{EQ:Binding_Energy_Grun} indicates that \citet{Grun_etal_1985} assumes $A_s \approx 300$ J kg$^{-1}$, which corresponds to a value for our model parameter of $\mathcal{P}_{A_s} = 3$. Our JFC model (both scenarios) requires $\mathcal{P}_{A_s} \approx 5000$. In other words, we require meteoroid binding energies  ${\sim}1000\times$ larger than those assumed in \citet{Grun_etal_1985}. 

\begin{figure}
\epsscale{1.2}
\plotone{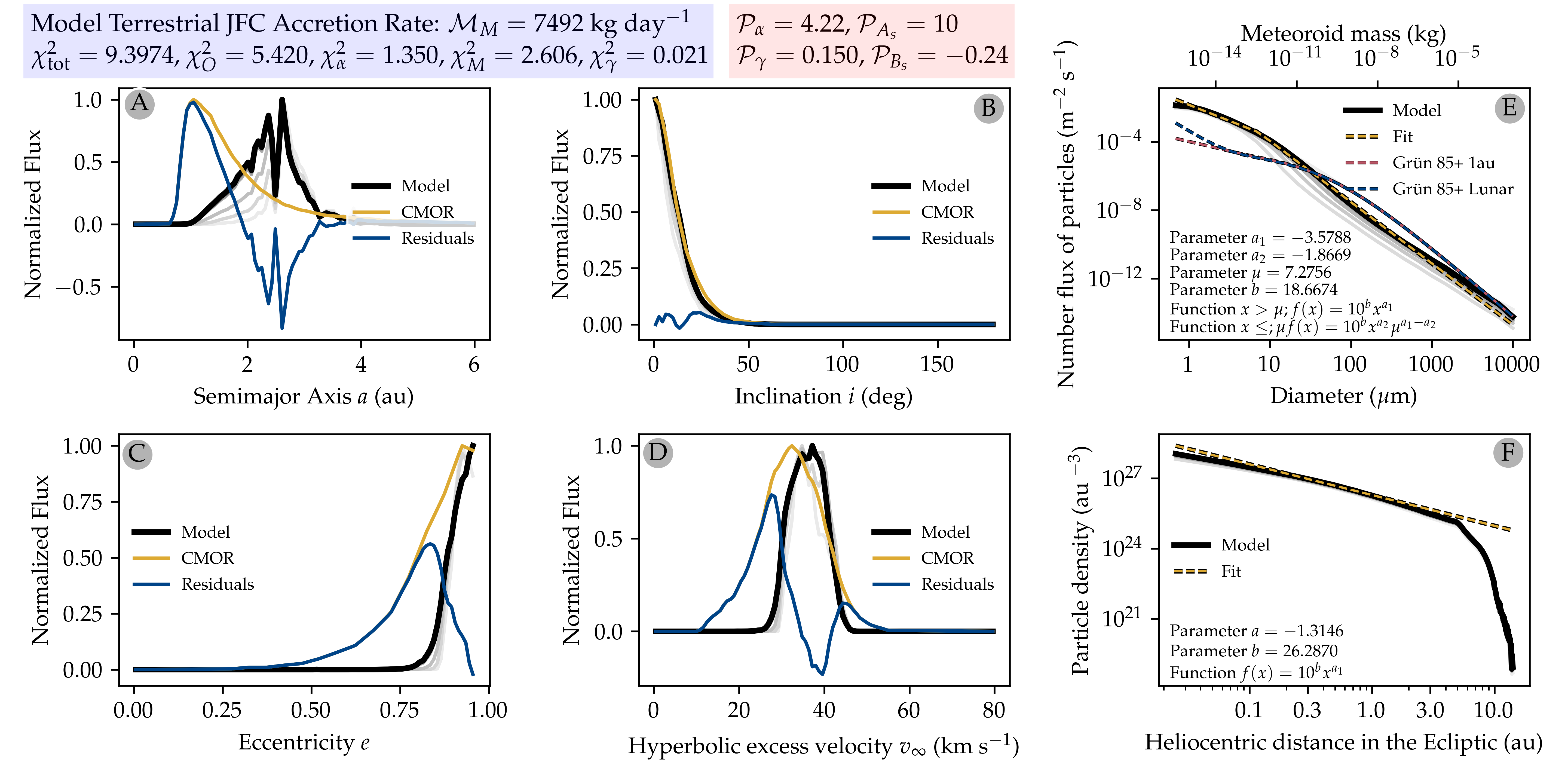}
\caption{\label{FIG:ScenarioB_Overview_Grun10}
The same as Fig. \ref{FIG:ScenarioA_Overview} but now for Scenario B assuming meteoroid binding energy close to that used in \citet{Grun_etal_1985}, i.e. $\mathcal{P}_{A_s} = 10$ ($A_s = 1000$ J kg$^{-1}$). Assuming much smaller $A_s$ than those providing best fit to our constraints leads to short dynamical lifetimes that prevent meteoroids detectable by CMOR to assume more circular orbits at semimajor axes close to 1 au. The size-frequency distribution at 1 au is also affected by removal of meteoroids with diameters $100 < D < 1000~\mu$m. The mass accretion rate is at 37\% of the best fit value in Fig. \ref{FIG:ScenarioA_Overview}. }
\end{figure}

Figure \ref{FIG:ScenarioB_Overview_Grun10} shows the model results assuming $\mathcal{P}_{A_s} \approx 10$, ($A_s = 1000$ J kg$^{-1}$), a value comparable to that used in \citet{Grun_etal_1985}. Since particles are much more prone to catastrophic destruction from much smaller particles than in our best fit model in both Scenarios, the larger meteoroids that are detectable by CMOR are efficiently removed before they can assume orbits with lower eccentricities and semimajor axis close to 1 au (Fig. \ref{FIG:ScenarioB_Overview_Grun10}A,C). The semimajor axis distribution of model meteors now peaks around $a=2.5$ au, i.e. inside the main-belt, which leads to large discrepancy with the meteor radar observation and low goodness-of-fit values. The eccentricities of meteoroids are kept close to those of the source population and cannot reproduce the abundance of radar meteors with $e<0.8$. This also impacts the observed meteor velocities, where the impacts with $v_\infty < 30$ km s$^{-1}$ are missing from the model. Subsequently, the size-frequency distribution at 1 au is affected, mostly for meteoroids with diameters $100 < D < 1000~\mu$m that are removed from the simulation before they can reach Earth. The larger meteoroids are not as affected because they do have shorter dynamical lifetimes due to their interaction with Jupiter \citep{Yang_Ishiguro_2018}. Frequent collisions also remove 63\% of the mass flux at Earth compared to the best model fit from Scenario B. 

\begin{figure}
\epsscale{1.2}
\plotone{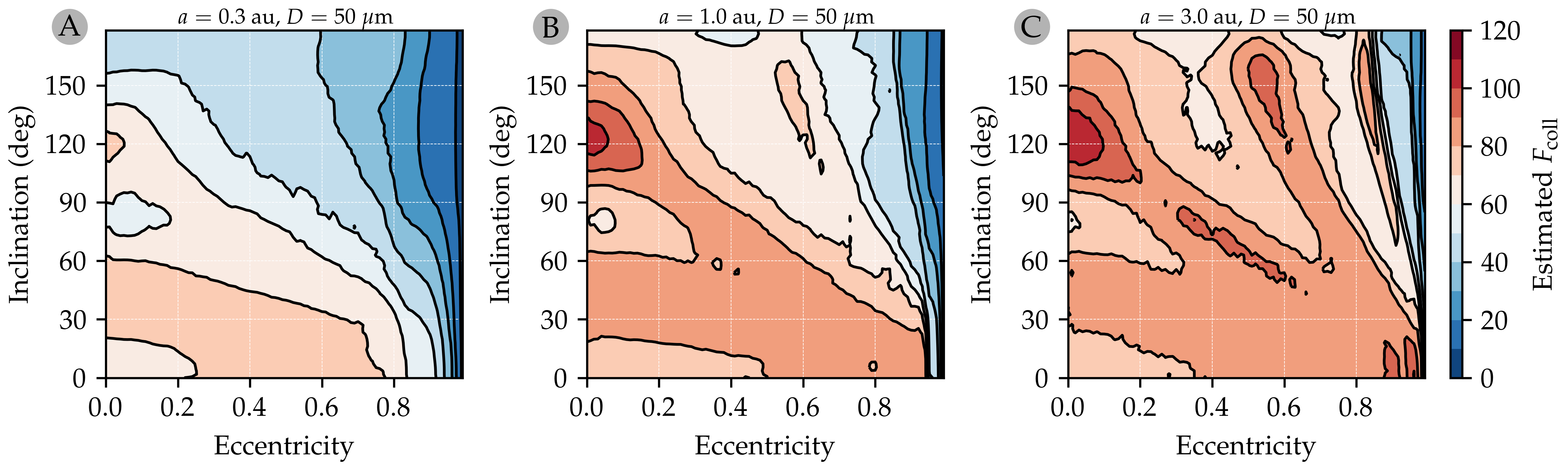}
\caption{\label{FIG:Collisions_Grun}
Estimated value of the collisional lifetime multiplier $F_\mathrm{coll}$ for a particle with $D=50~\mu$m and a full range of bound orbits phase space. We use \citet{Grun_etal_1985} binding energy estimate $A_s=300$ J kg$^{-1}$ and $A_s = 500,000$ J kg$^{-1}$ derived from our best fit model to derive the value of $F_\mathrm{coll}$. For the most of the orbital phase space the \citet{Grun_etal_1985} binding energy underestimates the collisional lifetimes by a factor of $40-100$. }
\end{figure}

This result is not new and surprising. \citet{Nesvorny_etal_2011JFC} showed that using the original \citet{Grun_etal_1985} assumptions leads to early removal of meteoroids from the simulation and the fit to radar data cannot be achieved. Similar conclusions have been shown in \citet{Soja_etal_2019}. These two works used a collisional lifetime multiplier, $F_\mathrm{coll}$, that increases the collisional lifetimes in the model by a factor of $F_\mathrm{coll}$. \citet{Nesvorny_etal_2011JFC} requires $F_\mathrm{coll}>30$ for JFC meteoroids to provide a good match to their constraints, while \citet{Soja_etal_2019} requires $F_\mathrm{coll}= 50$ for meteoroids with $D>250~\mu$m and $F_\mathrm{coll} = 1$ for smaller meteoroids. The reality is more complex as shown in Figure \ref{FIG:Collisions_Grun}, where we show the values of $F_\mathrm{coll}$ calculated from the ratio of the collisional lifetimes in our best model fit in Scenario B ($A_s = 500, 000$ J kg$^{-1}$) and approximate \citet{Grun_etal_1985} $A_s=1000$ J kg$^{-1}$ for three different values of semimajor axis $a$ and the full range of bound orbits in eccentricity and inclination assuming meteoroid diameter $D=50~\mu$m. In Figure \ref{FIG:Collisions_Grun}, we see that $F_\mathrm{coll}$ varies by an order of magnitude for particles on various orbits, where $F_\mathrm{coll}$ decreases with increasing eccentricity and inclination. In general, we can state that $30<F_\mathrm{coll}<100$ for orbits with $e<0.8$ for a wide range of semimajor axes in the inner solar system, which is in agreement with previous studies. However, seeing the complex picture resulting from our collisional grooming algorithm, we note that using simplistic estimates from analytically prescribed collisional lifetimes can under/over-estimate the collisional lifetimes by a factor of 3-5.

%%%%
%%%

%%%%
%%%

%%%%
%%%

\subsection{JFC mass production rate and total mass loss  rate and comparison to PSP}
\label{SEC:PSP_and_massLoss}
One of the output of the Parker Solar Probe voyage to the outskirts of the Sun was an estimate of $\beta$-meteoroid production rates and conversely the total mass production rate in the inner solar system \citep{Szalay_etal_2021}. In this Section, we calculate the mass loss our JFC dust cloud suffers in collisions, compare the mass loss and its scaling with the heliocentric distance to \citet{Szalay_etal_2021}, and estimate the JFC mass production rate needed to sustain the JFC portion of the Zodiacal Cloud.

To calculate the mass of dust that JFCs need to produce to sustain our best model fits in Scenarios A and B, we employ the following method. For each meteoroid diameter $D$ and initial pericenter $q^\star$ in our model, we calculate the total mass of meteoroids $M_\mathrm{nocoll}(D,q^\star)$ for a collision-less seed model and the collision-less dynamical lifetime $T_\mathrm{dyn}(D,q^\star)$. Using these two numbers we obtain the meteoroid production rate
\begin{equation}
    m_\mathrm{prod} = \sum_{D,q^\star}\frac{M_\mathrm{nocoll}(D,q^\star)}{T_\mathrm{dyn}(D,q^\star)}
\end{equation}

In Scenario A, we estimate $m_\mathrm{prod} = 9,360$ kg s$^{-1}$ while for Scenario B we obtain $m_\mathrm{prod} = 5,170$ kg s$^{-1}$. Both of these values agree with \citet{Nesvorny_etal_2011JFC} who estimated $10^{3} < m_\mathrm{prod}\mathrm{(JFC)} < 10^{4}$ kg s$^{-1}$. Our JFC production rates are approximately $5\times$ smaller than those reported in \citet{Yang_Ishiguro_2018} who estimated $29,000 < m_\mathrm{prod}\mathrm{(JFC)} < 53,000$ kg s$^{-1}$. \added{\citet{Rigley_Wyatt_2021} estimated the mass production rates of cometary fragmentation to the Zodiacal Cloud to have the mean value of $m_\mathrm{prod} = 990$ kg s$^{-1}$ and the median value of $m_\mathrm{prod} = 300$ kg s$^{-1}$ that are values an order of magnitude lower than our estimates. However, the same authors also find two epochs in their time dependent model where $m_\mathrm{prod} = 6,240$ kg s$^{-1}$ and $m_\mathrm{prod} = 11,110$ kg s$^{-1}$, which is very close to our values. Additionally, using the Helios spacecraft data, \citet{Leinert_etal_1983} found  $m_\mathrm{prod} =600-1000$ kg s$^{-1}$, which is more similar to the values found in \citet{Rigley_Wyatt_2021} and lower than our estimates.}

We can also calculate how much mass is produced in catastrophic collisions per unit of time by tracking the mass lost during each collision in our model. This is simply done by tracking the particle weights during the collisional grooming phase and multiplying by the mass that each particle in the model represents.
In our model the, catastrophic collisions remove $m_\mathrm{loss} = 430.6$ kg s$^{-1}$ and  $m_\mathrm{loss} = 164.5$ kg s$^{-1}$ for Scenarios A and B, respectively. This translates to 4.6\% and 3.2\% of \replaced{all}{the total mass of} particles in the simulation. These estimated mass losses are close to the $\sim$100 kg s$^{-1}$ estimated in \citet{Szalay_etal_2021} who used the $\beta$-meteoroid impact rates on the Parker Solar Probe to derive this number. The authors claim that 100 kg s$^{-1}$  is a lower limit due to their assumption that all collisional products are $\beta$-meteoroids. \replaced{This is unlikely, because}{This supports our findings since} larger meteoroids ($D>1$ mm) are likely to retain some mass in daughter products on bound orbits. Both of our scenarios allow a significant portion of mass lost in collisions to remain in bound orbits since our $m_\mathrm{loss}$ is higher than the one reported in \citet{Szalay_etal_2021}. \added{The bound daughter collisional products would be then subject to PR drag and radiation pressure and experience similar fate as the parent dust grains of their sizes.}

In contrast, \citet{Grun_etal_1985} estimate the $m_\mathrm{loss}=9000$ kg s$^{-1}$ for collisions inside 1 au. This value is similar to our meteoroid mass production rate that is needed to sustain the JFC meteoroid population in the Zodiacal Cloud. This would mean that all JFC particles in our model would be lost in collisions and no particles are lost to dynamical processes such as ejection of particles from the solar system, impacting one of the planets, and most importantly the particle disintegration near the Sun. \citet{Grun_etal_1985} estimates that 2.8\% of particles are lost to dynamical processes (PR drag) and the rest is lost to collisions. Our model predicts that 95.4\% and 96.8\% of mass is lost to dynamical processes for Scenarios A and B, respectively.

If we decrease the particle strength by a factor of ${\sim}1000$ to $A_s = 300$ J kg$^{-1}$ which we used to estimate the \citet{Grun_etal_1985} collisional lifetimes, the meteoroid loss rate in collisions is $m_\mathrm{loss}=1647$ kg s$^{-1}$ and $m_\mathrm{loss}=787$ kg s$^{-1}$ for Scenarios A and B, respectively. This translates to 17.6\% and 15.2\% of particles lost due to collisions for Scenarios A and B, respectively. Even in such a case, the majority of mass is lost to dynamical processes since especially smaller particles are either blown out of the solar system or destroyed by spiraling too close to the Sun. The difference of our results to  those reported in \citet{Grun_etal_1985} stems for a more comprehensive approach of our model by assuming the three-dimensional distribution of meteoroids in the Zodiacal Cloud and the relative impact velocities for each inter-particle collision. As we show in Figure \ref{FIG:Collisions_Comparison}, different combinations of orbital elements for particles with the same semimajor axis can lead to several orders of magnitude differences in collisional lifetimes. Also note that the $A_s = 300$ J kg$^{-1}$ model does not fit most of our observational constrains and our model requires values around $A_s = 500 000$ J kg$^{-1}$.

\begin{figure}
\epsscale{1.2}
\plotone{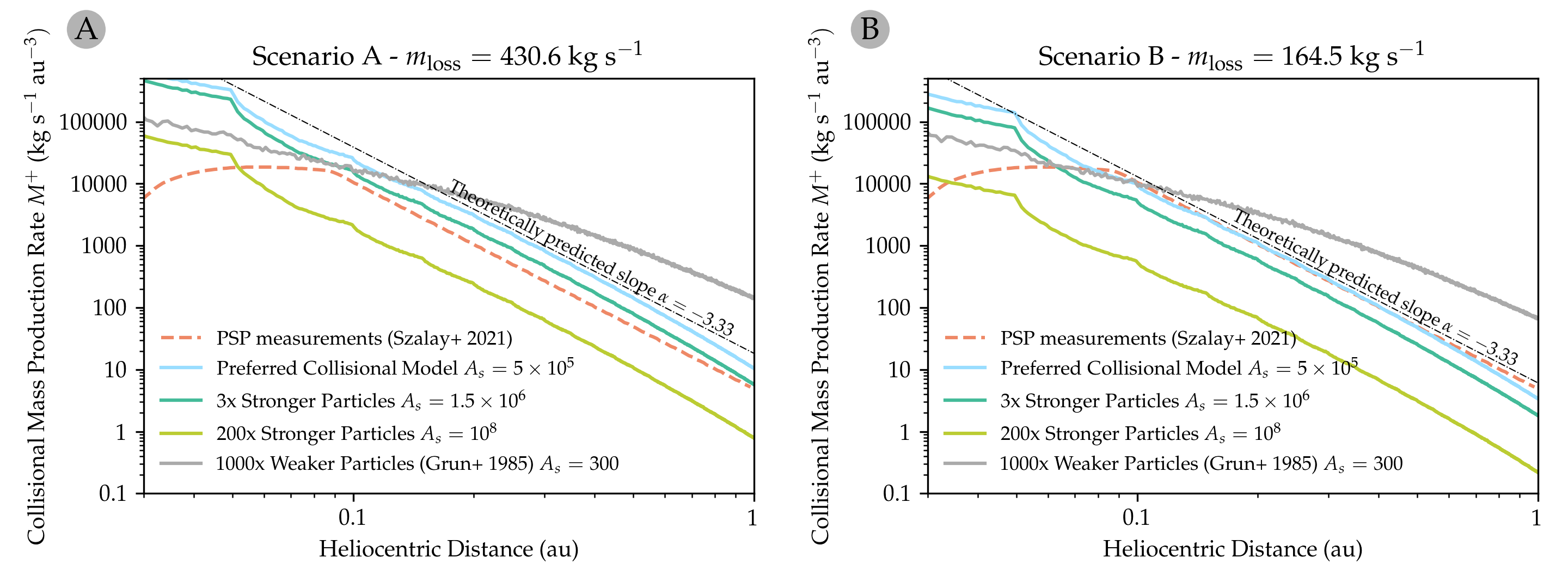}
\caption{\label{FIG:Radial_MassLoss}
\textit{Panel A}: The heliocentric distance scaling of the collisional mass production rate $M^+$ for Scenario A, where the total $m_\mathrm{loss}=430.6$ kg s$^{-1}$ for the best fit parameters. The dashed orange line shows the PSP measurements derived from \citet{Szalay_etal_2021}. The four realizations of our collisional model for JFCs are shown in color-coded solid lines: our preferred model $A_s=5\times10^5$ J kg$^{-1}$ (blue), $3\times$ stronger particles $A_s=1.5\times10^6$ (green) J kg$^{-1}$, $200\times$ stronger particles $A_s=10^8$ J kg$^{-1}$ (yellow), and $1000\times$ weaker particles $A_s=300$ J kg$^{-1}$ (grey). We also show the theoretically predicted scaling of $M^+$ based on brightness observations of the Zodiacal Cloud with a slope $\alpha=-3.33$ (dashed black line). Our preferred model shows a good match to PSP measurements as well as the predicted slope down to heliocentric distance of 0.1 au, where the effects of the dust depletion zone near the Sun are evident.
\textit{Panel B}: The same as Panel A but now for Scenario B, where the total mass loss is $m_\mathrm{loss}=164.5$ kg s$^{-1}$ for the best fit parameters. In Scenario B, we assumed a smaller total cross-section of the Zodiacal Cloud and thus all models give smaller values of $M^+$. Here, the $3\times$ stronger particle model with $A_s = 1.5\times10^6$ J kg$^{-1}$ underestimates $M^+$ compared to that derived from PSP measurements and provides an upper limit for $A_s$ in Scenario B.  }
\end{figure}

Our model also allows us to calculate the spatial distribution of mass produced in collisions per unit time, $M^+$. Figure \ref{FIG:Radial_MassLoss} shows the heliocentric distance scaling of $M^+$ for both scenarios A and B assuming four different values of $A_s$: our preferred model $A_s = 500, 000$ J kg$^{-1}$; a model with 3$\times$ stronger particles $A_s = 1, 500, 000$ J kg$^{-1}$; a model with 200$\times$ stronger particles $A_s = 10^8$ J kg$^{-1}$; and a model with ${\sim}1000\times$ weaker particles with $A_s = 300$ J kg$^{-1}$ \citep[the value used in][]{Grun_etal_1985}. Additionally, we compare our values to the \citet{Szalay_etal_2021} mass production rates derived from PSP observations, and the theoretically predicted scaling of $M^+$ with heliocentric distance $M^+ \propto r_\mathrm{hel}^{-3.3}$. To allow comparison with the \citet{Szalay_etal_2021} data, we adopt their equation for the total mass loss
\begin{equation}
    m_\mathrm{loss} =  \int_{0}^{2\pi} \int_{0}^{pi}\int_{r_0}^{r_1} M^+(r,\phi,\theta) r^2 \sin({\phi}) dr d\phi d\theta = 4\pi \varepsilon \int_{r_0}^{r_1} M^+(r) r^2 dr,
    \label{EQ:TotalMassLoss}
\end{equation}
where $r_0 = 0.025$ au and $r_1=1.0$ au, $\varepsilon=0.36$ is the filling factor that represents the latitudinal scaling of the mass production rate used in \citet{Szalay_etal_2021}, which follows the \citet{Leinert_etal_1981} conclusions. The second integral in Eq. \ref{EQ:TotalMassLoss} assumes perfect \replaced{radial}{azimuthal} symmetry for $M^+$ and the aforementioned filling factor. The PSP observations are limited to $r_1=1.0$ au, while our JFC dust cloud extends beyond Jupiter's orbit. However, the steep decrease in $M^+$ with heliocentric distance amounts to $<5\%$ difference if we extend $r_1 = \infty$ for our preferred model parameters.

For Scenario A, our model shows good agreement with the PSP measurements for our preferred model with $A_s = 500,000$ J kg$^{-1}$ as well as for particles $3\times$ stronger which provide $m_\mathrm{loss} =272.7$ kg s$^{-1}$. Much higher particle strength ($A_s=10^8$ J kg$^{-1}$) results in an order of magnitude smaller values of $M^+$ and $m_\mathrm{loss} =35.3$ kg s$^{-1}$. The slopes of our preferred model and models with higher particle strengths agree well with the theoretical slope estimated from the Zodiacal Cloud brightness scaling with heliocentric distance $M^+(r) \propto \mathcal{S}(r) r^{-1} \propto r^{-3.33}$, where $\mathcal{S}\propto r^{-2.33}\pm0.02$ \citep{Stenborg_etal_2018_Fcorona}. For much weaker particles ($A_s=300$ J kg$^{-1}$), our model predicts a different scaling with heliocentric distance where the bulk of the mass is produced beyond 1 au due to shorter collisional lifetimes of particles; $47\%$ of mass is lost in collisions beyond 1 au. For Scenario B, the values of $M^+$ are ${\sim}2.6\times$ smaller than in Scenario A, which moves our preferred model with $A_s = 500,000$ J kg$^{-1}$ very close to values from \citet{Szalay_etal_2021}. The slope of the heliocentric distance scaling of $M^+$ agrees well with the predicted slope and PSP measurements. 
For models with stronger particles, we get $M^+$ smaller than those estimated from PSP measurements ($m_\mathrm{loss} =92.3$ kg s$^{-1}$ for $A_s = 1,500,000$ J kg$^{-1}$) which provides an upper limit for the strength of particles in our model. For much weaker particles with $A_s = 300$ J kg$^{-1}$, we get similar results as in Scenario A: the slope of the $M^+$ scaling is significantly different from the observed value and $43\%$ of mass is lost in collisions beyond 1 au.

There is one aspect of the PSP measurement that our model cannot currently reproduce. \citet{Stenborg_etal_2021} reported that the dust depletion zone starts at a heliocentric distance of 0.1 au and continues inward. This results in a change of the slope of $M^+$ that can be seen in the PSP measurements in Figure \ref{FIG:Radial_MassLoss}. In our dynamical model, the JFC particles can survive inside 0.1 au which is reflected in an increase of $M^+$ in our collisionally groomed models. In order to capture this phenomenon, we would need to introduce an additional particle destruction mechanism that would clear the dust depletion zone and allowed us to match the plateau and decrease of $M^+$ inside 0.1 au. We leave this implementation for future work.

%%%%
%%%

\begin{figure}
\epsscale{1.2}
\plotone{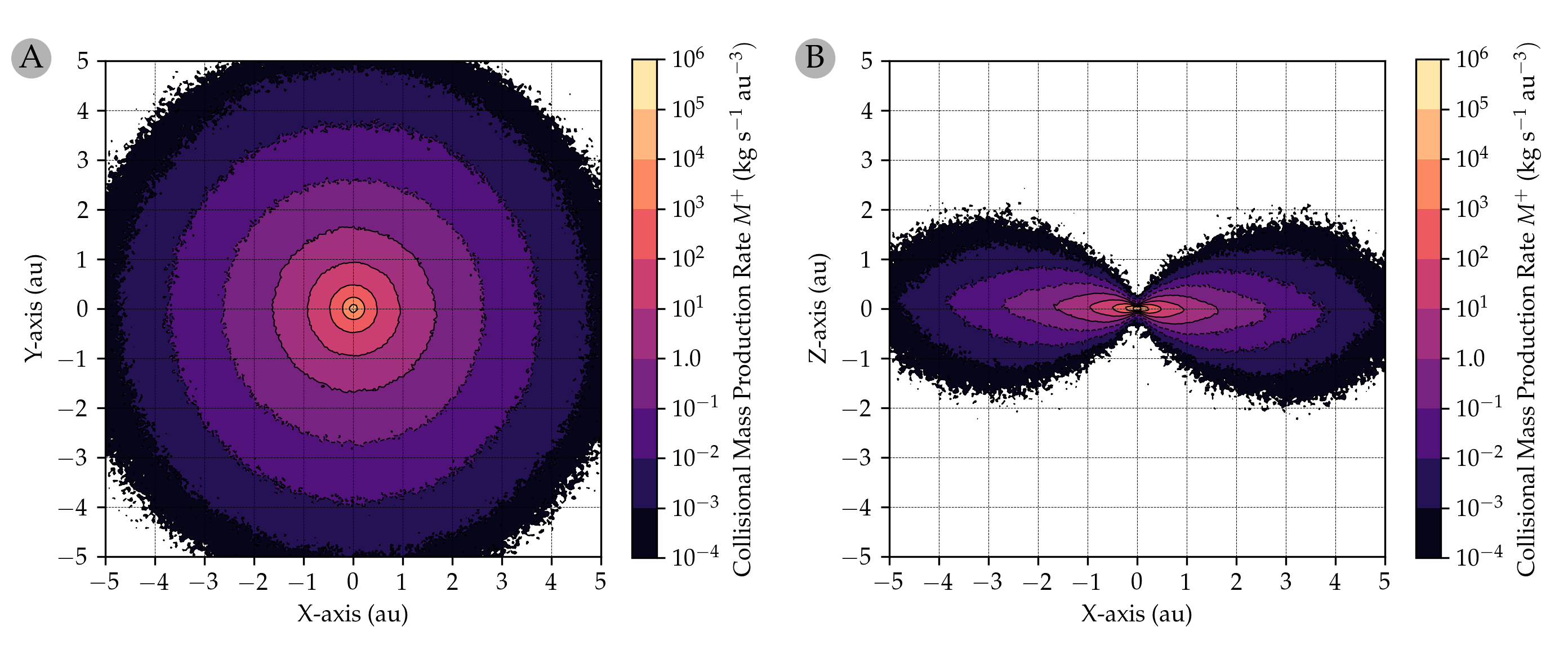}
\caption{\label{FIG:Spatial_MassLoss}
Panel A: Distribution of $M^+$ in the ecliptic plane in our preferred model in Scenario B. The plot shows the average $M^+$ within 0.025 au from the ecliptic plane. The value of $M^+$ in the ecliptic is radially symmetric in our model and decreases as $M^+\propto r_\mathrm{hel}^{-3.83}$ between $0.1 < r_\mathrm{hel} < 1.0$ au, which is steeper than the radially averaged value of $M^+$ shown in Fig. \ref{FIG:Radial_MassLoss}. The color-coded contours show the order of magnitude increase in $M^+$ where our model predicts 9 orders of magnitude difference between $M^+$ at $r_\mathrm{hel}=5$ au and $r_\mathrm{hel}=0.1$ au.
Panel B: The same as Panel A but now for a slice through the $Y$-axis, a side view of the JFC dust cloud where the values of $M^+$ are averaged over 0.025 au away from the $Y$-axis in our model. Due to rotational symmetry these two images can be used to reconstruct the entire 3D image of the distribution of $M^+$ in our JFC model.}
\end{figure}

The radial scaling of averaged values of $M^+$ does not provide a complete picture of the collisional processing of the Zodiacal Cloud. In Figure \ref{FIG:Spatial_MassLoss}, we show two slices through our model: slice through the $Z-$axis ($X-Y$ or ecliptic plane; Figure \ref{FIG:Spatial_MassLoss}A and through the $Y$ axis ($X-Z$ plane, Figure \ref{FIG:Spatial_MassLoss}B).  For both slices, we average the value of $M^+$ over 0.025 au away from the slice-through axis. The distribution of $M^+$ in the ecliptic plane is azimuthally symmetric as expected from our JFC simulation setup where we record particle and planet positions every 100 years and planets are sampled at random positions along their orbits  (Fig. \ref{FIG:Spatial_MassLoss}A). We can also calculate the scaling of $M^+$ with heliocentric distance, where get a value of $M^+\propto r_\mathrm{hel}^{-3.83}$ between $0.1 < r_\mathrm{hel} < 1.0$ au with a steeper decrease beyond 1 au $M^+\propto r_\mathrm{hel}^{-5.5}$. The slope of the $M^+$ scaling in the ecliptic is steeper (by 0.5 units) than the value for the volume average shown in Fig. \ref{FIG:Radial_MassLoss}. This is due to higher values of $M^+$ as well as the particle density close to the ecliptic, which is obvious from the side view shown in Fig. \ref{FIG:Spatial_MassLoss}B. We fit the latitudinal profiles of $M^+$ using a function $M^+(\beta) = A \exp(-B|z/r_\mathrm{hel}|)$ and obtained the value of the fitting parameter $B=8.3\pm0.3$, which is smaller than $B=2.1$ derived by \citet{Leinert_etal_1981} from the Helios observations of the brightness of the inner Zodiacal Cloud. This shows that JFC meteoroids provide the close to the ecliptic component and the more diffuse component from Halley-type and Oort Cloud comets is needed to capture the entire image of the Zodiacal Cloud.

%%%%
%%%

\section{Consequences for solar and exo-solar dust and debris clouds}
\label{SEC:EXOSOLAR}
Our models show that the collisional strength of particles originating from Jupiter-family comets is ${\sim}3$ orders of magnitude higher than values commonly assumed in various solar system and debris disk studies. For dust originating from the Edgeworth-Kuiper Belt (EKB) objects, \citet{Kuchner_Stark_2010} assumed $A_s = 20$ J kg$^{-1}$, whereas in \citet{Poppe_2016} and \citet{Poppe_etal_2019} the authors assumed that any collision results in a catastrophic disruption; i.e $A_s = 0$. These values are $>25,000\times$ smaller than those we derived from our models and are even smaller than those assumed in \citet{Grun_etal_1985}. It is understandable that EKB particles have a significant icy component that is likely to be less resistant to collisions and our JFC model cannot test that. The impact experiments of ice-silicate mixtures show that the higher percentage of ice content leads to higher impact yield and thus weaker particle binding energy \citep{Koschny_Grun_2001}. However, these results do not show 4 orders of magnitude changes in any measured quantity. Moreover, \citet{Benz_Asphaug_1999} show in their work that basalt and ice have similar collisional strength.  For these reasons, we think that the effects of collisional grooming in \citet{Kuchner_Stark_2010}, \citet{Poppe_2016}, and \citet{Poppe_2019} are overestimated and should be revisited with a particle setup more resilient to collisions. As we show in Sec. \ref{SEC:PSP_and_massLoss} the value of $A_s$ strongly influences the mass lost in collisions and the production of $\beta$-meteoroids, where more resilient particles can lead to 1-2 orders of magnitude smaller mass losses. This might lead to overestimation of the mass produced in collisions, shortening of the dynamical lifetime of particles in the model that are not able to get close to the host star, as well as underestimating the total mass of the dust source population if the total brightness of the dust cloud is used to constrain it.

% Several exo-solar debris disk  studies used formalism and parameters that we can compare to the results of this work.
% \citet{Lohne_etal_2012} uses a nominal value of $A_s = 500$, which is $1000\times$ smaller than values we obtained from our collisionally processed model JFC particles. Since all JFC meteoroid orbits reach below the snow line \citep[$\lesssim 3$ au,][]{Jewitt_etal_2007} we can assume that the icy content of JFC meteoroids is negligible.  In their model of HD 207129 \citet{Lohne_etal_2012} found that assuming higher content of ice in their particles ($>50\%$) provides better fit to the observation than models with purely silicate dust grains. \citet{Lohne_etal_2012} tested the effects of increased and decreased value of $A_s$ by a factor of 3. They concluded that higher or lower values of $A_s$ do not provide as good fit to the observation as their preferred value.

Most exo-solar debris disk observations are modeled using three different codes that estimate the effects of collisions: ACE \citep{Krivov_etal_2005}, DyCoSS \citep{Thebault_etal_2012}, and LIDT-DD \citep{Kral_etal_2013}. We find that the values of $A_s$ or the critical specific energy $Q^*_D$ used in models using these algorithms are mostly derived from the \citet{Benz_Asphaug_1999} work or are assumed to be smaller than in \citet{Benz_Asphaug_1999}. In \citet{Benz_Asphaug_1999} the authors used smooth particle hydrodynamics (SPH) method to simulate collisions of cm-to-km sized bodies using basaltic and icy particles, where the impact velocities were considered to be between 0.5 km s$^{-1}$ and 5.0 km s$^{-1}$. For basaltic material, the authors find $A_s = 9000$ J kg$^{-1}$ and $A_s = 3500$ J kg$^{-1}$ for impact speeds of 5 km s$^{-1}$ and 3 km s$^{-1}$ respectively. For icy material the values are: $A_s = 1600$ J kg$^{-1}$ and $A_s = 7000$ J kg$^{-1}$ for impact velocities 3 km s$^{-1}$ and 0.5 km s$^{-1}$, respectively. In general, these values are $\sim100\times$ smaller than those derived in our work. 
Unfortunately, the \citet{Benz_Asphaug_1999} work does not discuss effects on particles smaller than 1 cm in diameter. This might be a potential source of discrepancy between our results and the values reported in \citet{Benz_Asphaug_1999} that are used in many debris disk studies. Such smaller values of $A_s$ lead, in general, to shorter dynamical lifetimes of all meteoroids in debris disk systems, higher mass production in collisions, and underestimation of total dust mass present in debris disks. However, each exo-solar system is different and the particular consequences of smaller/larger particle critical specific energy $Q^*_D$, that is linearly proportional to $A_s$, need to be resolved by extensive modeling efforts that are beyond the scope of this article. 

\section{Discussion}
We provided the first implementation of a dynamical model for JFC meteoroids that includes iterative collisional grooming and which fits multiple observational constraints available in the inner solar system. Some aspects of the model require further improvement. We discuss these aspects here.

In this work, we assume that JFC particles dominate the inner solar system Zodiacal Cloud in terms of particle density, cross-section, and mass. Many studies support this picture \citep{Nesvorny_etal_2010,Poppe_2016,Soja_etal_2019, Yang_Ishiguro_2018} and our work shows that JFC dust production can well reproduce all constraints we present in this article. By adding additional major dust populations such as main belt asteroid meteoroids, Halley-type comet meteoroids, and Oort cloud comet meteoroids, we do not expect significant changes in the best parameter fit we find for our JFC model. However, we expect that addition of particles with a broader inclination range, including some on retrograde orbits, will increase the rates of collisions in the innermost regions of the solar system because of how the particle density scales with the heliocentric distance \citep{Poppe_2016, Pokorny_Kuchner_2019}. 

%Our process of finding the best JFC model fit to our constraints examined two scenarios that treated two particle clouds differing in the total particle cross-section by a factor of two. 
During our search for the best fit of our JFC meteoroid model to our set of constraints we explored two different scenarios with differing total particle cross-sections: (A) $\Sigma_\mathrm{ZC} = 2.0 \pm 0.5 \times 10^{17}$ m$^{2}$ based on \citet{Nesvorny_etal_2011JFC}, and (B)  $\Sigma_\mathrm{ZC} = 1.12 \times 10^{17}$ m$^{2}$ based on \citet{Gaidos_1999}.
We showed that we were able to reconcile the Scenario A fit by using the mass accretion rate of JFCs at Earth from \citet{CarrilloSanchez_etal_2016}, whereas in Scenario B we show that the \citet{Gaidos_1999} estimate of the total Zodiacal Cloud cross-section matches well the mass accretion rates from \citet{CarrilloSanchez_etal_2020}. We conclude that more constraints are needed to decide which of the two presented scenarios truly represents the current state of the Zodiacal Cloud. 

Our collisional grooming model itself is still an approximation of reality where we assume that meteoroid collisions have either no effect or they catastrophically disrupt particles involved in the collision. \added{We also assume that the dust produced in collisions does not further influence the collisional evolution of dust in our model and that all collisional fragments are beta-meteoroids.} We also assume that the Zodiacal Cloud and our Jupiter-family comet meteoroid model is unchanging in time and is averaged over timescales longer than orbital periods of the eight planets. This is not entirely correct since observations \added{and models} show that the Zodiacal Cloud experiences time variability \added{on thousand to million year timescales} \citep{Napier_2001,Koschny_etal_2019, Rigley_Wyatt_2021}.

While we attempted to acquire a very broad selection of model constraints, we acknowledge that more data sets are available that were not discussed in detail in this article. We are seeking to adopt an even broader set of observations and measurements that would allow us to better constrain the meteoroid production rates in collisions, provide independent measurements for mass accretion rates at various planets, and set constraints on the global shape of the Zodiacal Cloud for a broad range of dust and meteoroid sizes. With our model available to the general public, we aim to keep the model updated via community updates on the project website.

One of the potential drawbacks of our current model for JFC dust population is the assumption of a single power law of particles generated by the source population. We decided to avoid a more complicated SFDs such as the broken power law due to the additional fitting parameters (2 for the broken power law). This increase in the size of the parameter space would significantly extend the total processing time required to find the best fits to our suite of constraints. Introducing a broken power law would influence the particle number fluxes shown in Figs. \ref{FIG:ScenarioA_Overview} and \ref{FIG:ScenarioB_Overview} as well as the particle density with respect to heliocentric distance. It would also change the collisional lifetimes of particles with diameters close to and below the break point of the power law. We do not expect significant changes for the strength of the particles $A_s$ derived in this work since the most collisionally impacted are larger particles with $D>100~\mu$m and we expect the break points to be around $D\sim30~\mu$m based on our initial analysis and the FIRAS measurements of the Zodiacal Cloud spectrum by \citet{Fixsen_Dwek_2002}.

\subsection{Time and memory consumption of the collisional grooming process}
The collisional grooming process run time is influenced by two major factors: (1) the total number of records in the model/cloud and (2) the number of iterations needed to attain a steady state. Most of our modeling runs took 5-6 iterations to reach the first steady state and then 1-2 further iterations to reach the second, rescaled steady state. The iteration time varied between different computer setups, ranging between 3000 and 4000 seconds per iteration, or approximately 6-8 hours for one model run. 

The original memory required per run was over 128GB RAM which limits the model deployment on most computers available to us.
%This might limit deployment on machines with RAM limitations, but should not present a problem for most modern computers. 
Therefore, we implemented several tweaks that reduce the memory requirements without decreasing the precision of the collisional grooming method.
This was done through the aggregation of particles inside each volume element that had the same diameter and similar velocity vectors. The particle position information is not retained once the particle is assigned to a particular volume element, thus it plays no role in particle aggregation.  The velocity vector similarity is checked in two steps: first the angular difference $\vec{v_1}\cdot \vec{v_2}/(\| v_1 \|\| v_2 \|) = \cos{\phi}$ is checked, and then the velocity vector size ratio is checked $R_v = |\log \| v_1 \| - \log\| v_2 \||$. In our model, we aggregate particles with the same $D$ that have a velocity vector angular difference $<10^\circ$ ($\cos{\phi} > 0.985)$ and their vector size ratio is $<10\%$ ($R_v <0.0414)$. Making this approximation reduced the memory usage to approximately 17.3~GB of RAM. This memory requirement still limits deployment on base-model laptops, which usually have 8-16 GB of RAM, but it suits the capabilities of a typical ``professional" desktop computer with 32+ GB of RAM. We tested the loss of precision by running the untweaked version of the code and compared the results with our particle aggregation optimization. The difference in $\chi^2$ values in both models was $<1\%$.

% \section{Future work}
% In the future work we aim to add the following aspects:
% \begin{itemize}
%     \item Add multi-CPU support and allow efficient parallel computing
%     \item Add GPU support to benefit from parallel processing
%     \item Add an adaptive grid to handle better the overdensities in the particle clouds (e.g., close to the central star)
%     \item Include more constraints in our analysis to make the Zodiacal Cloud model more reliable
%     \item Add other significant dust and meteoroid producing populations such as the main-belt asteroids \citep{Nesvorny_etal_2006} and long-period comets \citep{Nesvorny_etal_2011OCC, Pokorny_etal_2014}
% \end{itemize}

\section{Conclusions}
We developed a new code for collisional grooming of any extensive dust/particle cloud. Results of this work are primarily aimed at the main component of the Zodiacal Cloud - the JFCs; however, as we show in the Validation section, our code can handle virtually any distribution of dust and it is only limited by the amount of memory and processing power.

Here, we summarize the most important findings:
\begin{itemize}
    \item We developed a new code following the algorithm and methods from \citet{Stark_Kuchner_2009} that is available at our project page. A simple planet-less test case agrees with an analytic solution for various dust cloud optical depths (Fig. \ref{FIG:Simple_Stark_Test})
    \item Our Jupiter-family comet meteoroid model provides a satisfactory match to all the observational constraints we investigated: the heliocentric distance density distribution of the Zodiacal Cloud, the mass accretion rate of Jupiter-family Comet meteoroids at Earth, the orbital distributions of radar meteors, and the size-frequency distribution of meteoroids at 1 au.
    \item The best model fit was found for the following parameters: the differential size-frequency index at the source $\alpha = -4.20 \pm 0.10$, the material constant $A_s = 5\times 10^5 \pm 4\times 10^5$ J kg$^{-1}$, $B_s = -0.24$, and $\gamma = 0.20 \pm 0.025$
    \item From our model, we expect Jupiter-family comets to eject meteoroids and dust with a differential size-frequency index $\alpha = -4.20 \pm 0.10$ following the single-power law. 
    \item The collisional lifetime multiplier used in previous dynamical models does not characterize the collisional lifetimes predicted by our model (Fig. \ref{FIG:Collisions_Comparison})
    \item We find that meteoroids have approximately $1000\times$ larger binding energy than suggested in \citet{Grun_etal_1985} and \citet{Krivov_etal_2005} and $100\times$ larger than values from \citet{Benz_Asphaug_1999}. This difference might have a significant impact on exo-solar debris disk studies that include inter-particle collisions. Such models would underestimate the effect of PR drag, overestimate the mass produced in collisions, and underestimate the mass of the dust source population.
    \item Our Jupiter-family comet meteoroid model requires $9360$ kg s$^{-1}$ or $5170$ kg s$^{-1}$  of material produced to sustain the steady state for Scenarios A and B, respectively. We estimate that around 3-5\% of such mass is lost in catastrophic collisions and the rest is lost in dynamical processes such as PR drag, accretion by planets, and the ejection from the solar system.
    \item We find that our models match well the Parker Solar Probe heliocentric distance profile of the \added{collisional} mass production rates \added{of $\beta$-meteoroids} \citep{Szalay_etal_2021} (Fig. \ref{FIG:Radial_MassLoss}).
\end{itemize}

\acknowledgements
{
Support for this research was provided by NASA’s Planetary Science Division Research Program, through ISFM work packages EIMM and Planetary Geodesy at NASA Goddard Space Flight Center, NASA award numbers 80GSFC21M0002, \added{80NSSC21K1764,} and 80NSSC21K0153, and Grant Agency of the Czech Republic, grant number: 20-10907S. 
}

\software{
GitHub repository ({\url{ https://github.com/AnonymizedForPeerReview}})
$\bullet$ \texttt{gnuplot} (\url{http://www.gnuplot.info}) $\bullet$
\texttt{swift} \citep{Levison_Duncan_2013} $\bullet$ \texttt{matplotlib} (\url{https://www.matplotlib.org}) \citep{Hunter_2007} $\bullet$
\texttt{SciPy} (\url{https://www.scipy.org})
}

\bibliography{Papers}{}
\bibliographystyle{aasjournal}

\end{document}